\documentclass{aastex63}%[twocolumn]
\usepackage{lineno}
\usepackage{amsmath}
\usepackage{mathrsfs}

\submitjournal{ApJL}
\shorttitle{Baryon Loading in GRB~170817A}
\shortauthors{Ren et al.}
\begin{document}
\title{Constraining the Jet Launching Time of GRB 170817A by Utilizing the Baryon Loading}

\correspondingauthor{Da-Bin Lin}
\email{lindabin@gxu.edu.cn}

\author[0000-0002-9037-8642]{Jia Ren}
\author[0000-0001-6624-6052]{Da-Bin Lin}
\author{Lu-lu Zhang}
\author{Kai Wang}
\author{Xiao-yan Li}
\author[0000-0001-8411-8011]{Xiang-Gao Wang}
\author[0000-0002-7044-733X]{En-Wei Liang}
\affiliation{Laboratory for Relativistic Astrophysics, Department of Physics, Guangxi University, Nanning 530004, China}

\begin{abstract}
The observed delay of GRB~170817A relative to GW170817
carries significant information about gamma-ray burst (GRB) physics
and is subject to intense debate.
In this letter,
we put forward an approach to discuss the major source of this time delay.
First of all, we use the structured jet model to fit
the X-ray/optical/radio afterglows of GRB~170817A
together with superluminal motion measured by
the Very Long Baseline Interferometry.
Our structured jet is modelled with
angle-dependent energy and baryon loading.
It is found that our model can well fit the afterglows of GRB~170817A.
After that, the baryon loading in the jet is inferred
based on our fitting results.
By comparing the baryon loading to
the mass outflow in different stages,
we infer that the time lag of the jet launch relative to the merger
is less than hundreds or tens of milliseconds.
It suggests that the time delay of GRB~170817A relative to GW170817
is defined mostly by the spreading time
of the jet propagating to its dissipation radius.
\end{abstract}
%%%%%%%%%%%%%%%%%%%%%%%%%%%%%%%%%%%%%%%%%%%%%%%%%%%%%%%%%%%%%%%%%%%%%%%%%%%%%%%%%%%%
%%%%%%%%%%%%%%%%%%%%%%%%%%%%%%%%%%%%%%%%%%%%%%%%%%%%%%%%%%%%%%%%%%%%%%%%%%%%%%%%%%%%
\section{Introduction}

On August 17, 2017 at 12:41:04 UTC,
the Advanced Laser Interferometer Gravitational-wave Observatory and the
Advanced Virgo gravitational-wave detectors made their first observation
of a gravitational wave (GW) event, GW170817, from binary neutron star (NS) merger
(\citealp{Abbott_BP-2017-Abbott_R-ApJ.850L.39A,
Abbott_BP-2017-Abbott_R-ApJ.850L.40A,
Abbott_BP-2017-Abbott_R-ApJ.841.89A,
Abbott_BP-2017-Abbott_R-PhRvD.96b2001A}).
GW170817 was followed by a short gamma-ray burst (GRB), GRB~170817A
(\citealp{Abbott_BP-2017-Abbott_R-ApJ.848L.13A,
Goldstein_A-2017-Veres_P-ApJ.848L.14G,
Zhang_BB-2018-Zhang_B-NatCo.9.447Z}),
which triggered the Fermi Gamma-ray Burst Monitor at $t_{\rm obs}\sim 1.7$~s after the GW signal
and lasted for $\sim 2$~s.
The delay of GRB~170817A relative to GW170817 is subject to intense debate on the field of GRBs
(\citealp{Zhang_B-2018-pgrb.book.....Z,
Zhang_B-2019-FrPhy.1464402Z, Burns_E-2019arXiv190906085B}).
Except for GRB~170817A,
recent controversial gamma-ray signals,
GBM-150914
(\citealp{Connaughton_V-2016-Burns_E-ApJ.826L.6C,
Greiner_J-2016-Burgess_JM-ApJ.827L.38G, Connaughton_V-2018-Burns_E-ApJ.853L.9C})
and GBM-190816
(\citealp{Yang_YS-2019-Zhong_SQ-arXiv191200375Y}),
were claimed to follow the black hole - black hole (BH-BH) or BH-NS merger GW signals,
GW150914 and GW190816,
with a lag of $\sim 0.4$~s and $\sim 1.57$~s, respectively.
It indicates that the time delay of GRB relative to GW signal
may be common in compact binary mergers.
The origin of the time delay for GRB with respect to the GW signal has two arguments mainly.
Some authors attributed to the co-effect of
the delayed jet launching
and the jet breakout from the ejecta
(e.g., \citealp{Gottlieb_O-2018-Nakar_E-MNRAS.479.588G,
Bromberg_O-2018-Tchekhovskoy_A-MNRAS.475.2971B}).
Other authors (e.g., \citealp{Zhang_BB-2018-Zhang_B-NatCo.9.447Z, Lin_DB-2018-Liu_T-ApJ.856.90L})
suggested that the jet may be launched promptly after the merger
and the delay is mostly defined by the spreading time
when the jet propagates to the dissipation radius.
However, there is no consensus so far.

The binary NS mergers are expected to release an amount of neutron-rich
matter (\citealp{Lattimer_JM-1974-Schramm_DN-ApJ.192L.145L, Lattimer_JM-1976-Schramm_DN-ApJ.210.549L, Symbalisty_E-1982-Schramm_DN-ApL.22.143S}),
which can synthesize the elements heavier
than iron via the rapid neutron-capture process ($r$-process).
Numerical simulations of NS-NS mergers reveal that
there has a large number of matter outflowing from the system
(see \citealp{Nakar_E-2019-arXiv191205659N} for a detailed introduction),
e.g.,
the dynamic ejecta is formed in the first $\sim 10$~ms after the merger;
the neutrino-driven wind is stripped from the accretion disk and central object;
the viscosity-driven wind is blown off by the disk heating.
(Hereafter, the dynamic ejecta, the neutrino-driven wind, and the viscosity-driven wind
are all named as ``outflow'' in order to be distinguished with the ultra-relativistic jet.)
Owing to the nucleosynthesis,
there is a lot of radioactive heavy elements in the outflowing material.
Correspondingly, a kilonova powered by the radioactive decay of heavy elements will appear
(\citealp{Li_LX-1998-Paczynski_B-ApJ.507L.59L}; also see \citealp{Metzger_BD-2017LRR.20.3M} for a review),
e.g., AT2017gfo (\citealp{Coulter_DA-2017-Foley_RJ-Sci.358.1556C}).
In different post-merger stages, the outflow has different properties,
e.g., mass, angular distribution, and electron abundance, which would affect the outcomes of synthesized elements.
Thus, the ``red'', ``blue'', and even ``purple'' component emerges in the observations of kilonova
(e.g., \citealp{Villar_VA-2017-Guillochon_J-ApJ.851L.21V, Zhu_JP-2020-Yang_YP-ApJ.897.20Z}).
Based on this multi-components prescription,
plenty of works presented their estimation on the ejecta properties by modelling and fitting AT2017gfo
(e.g., \citealp{Kasen_D-2017-Metzger_B-Natur.551.80K, Villar_VA-2017-Guillochon_J-ApJ.851L.21V,
Cowperthwaite_PS-2017-Berger_E-ApJ.848L.17C, Waxman_E-2018-Ofek_EO-MNRAS.481.3423W}).
Furthermore, the possible effect of the compact remnant on AT2017gfo has been also studied by some groups
(e.g., \citealp{Yu_YW-2018-Liu_LD-ApJ.861..114Y, Ren_J-2019-Lin_DB-ApJ.885.60R,
Li_SZ-2018-Liu_LD-ApJ.861L.12L, Matsumoto_T-2018-Ioka_K-ApJ.861.55M}).
It is suggested that the compact remnant at least survived as an NS for some time,
thereby blown out enough material to power AT2017gfo (e.g., \citealp{Metzger_BD-2018-Thompson_TA-ApJ.856.101M}).

When an energetic jet expands outward,
it will be unavoidable to be contaminated by baryons in the outflow.
Thus, we propose the following idea.
With a relatively detailed understanding
of the spatial and temporal distribution of the merger outflows,
whether it is possible to infer the waiting time of jet launching
by comparing the baryon loading of the jet with the outflows in different post-merger stages?
In this {\em Letter},
we perform fitting of GRB~170817A afterglow
to infer the baryon loading of the jet in GRB~170817A
and further discuss the jet launching time.
The paper is organized as follows.
In Section~2, we introduce the models and methods used in our fitting.
In Section~3, we give the fitting result and the corresponding discussion.
The summary is made in Section~4.

%%%%%%%%%%%%%%%%%%%%%%%%%%%%%%%%%%%%%%%%%%%%%%%%%%%%%%%%%%%%%%%%%%%%%%%%%%%%%%%%%%%%%%%%%%%%%%%%%%%%%%%%%%%%%%%%
%%%%%%%%%%%%%%%%%%%%%%%%%%%%%%%%%%%%%%%%%%%%%%%%%%%%%%%%%%%%%%%%%%%%%%%%%%%%%%%%%%%%%%%%%%%%%%%%%%%%%%%%%%%%%%%%
\section{Model and Superluminal Motion}
To describe the jet structure and the dynamics of the external-forward shock,
we introduce a spherical coordinate ($r, \theta, \varphi$) with $r=0$ locating at the burst's central
engine and $\theta=0$ being along the jet axis.
We assume the observer locating at the direction of ($\theta_{\rm v}$, $\varphi_{\rm v}$) with $\varphi_{\rm v}=0$.

\emph{\textbf{Structured jet Description}\;}
Different from the previous works,
the structured jet is modelled with the angle-dependent baryon loading $\rho(\theta)$
and kinetic energy $\varepsilon(\theta)$ per solid angle in this work.
We consider an axisymmetric power-law structured jet, i.e.,
\begin{equation}
E_{\rm{iso}}(\theta ) = E_{\rm{iso,0}}
{\left( {1 + \frac{\theta }{{{\theta _\varepsilon}}}} \right)^k},
\end{equation}
\begin{equation}
M_{\rm{iso}}(\theta ) = M_{\rm{iso,0}}{\left( {1 + \frac{\theta }{{{\theta _\rho}}}} \right)^s},
\end{equation}
and an axisymmetric Gaussian structured jet, i.e.,
\begin{equation}
E_{\rm{iso}}(\theta ) = E_{\rm{iso,0}}\exp{\left( - \frac{\theta^2}{2\theta _\varepsilon^2} \right)},
\end{equation}
\begin{equation}
M_{\rm{iso}}(\theta ) = M_{\rm{iso,0}}{\exp{\left( - \frac{\theta^2}{2\theta _\rho^2} \right)}},
\end{equation}
where $E_{\rm iso}(\theta)=4\pi\varepsilon(\theta)$,
$M_{\rm iso}(\theta)=4\pi\rho(\theta)$,
and
$\theta_\varepsilon$ and $\theta_\rho$ are the characteristic half opening angle of
$\epsilon(\theta)$ and $\rho(\theta)$, respectively.

\textbf{\emph{Dynamics of the external-forward shock}\;}
The hemisphere which centers in the jet axis
is divided into $400 \times 100$ small patches
along $\theta$ and $\varphi$ directions in their linear space.
In this work, we consider the jet has no lateral expansion (but see \citealp{Troja_E-2019-vanEerten_H-MNRAS.tmp.2169T}),
and the dynamics of the external-forward shock is estimated independently in each patches,
i.e. (\citealp{Zhang_B-2018-pgrb.book.....Z}),
\begin{equation}\label{eq:dGamma}
\frac{d\Gamma}{dr}=
-\frac{\Gamma(\Gamma^2-1)(\hat{\gamma}\Gamma-\hat{\gamma}+1)\frac{dm}{dr}c^2
-(\hat{\gamma}-1)\Gamma(\hat{\gamma}\Gamma^2-\hat{\gamma}+1)(3U/r)}
{\Gamma^2[\rho(\theta)+m]c^2+(\hat{\gamma}^2\Gamma^2-\hat{\gamma}^2+3\hat{\gamma}-2)U}~,
\end{equation}
\begin{equation}\label{eq:diff_U}
 \frac{d U}{d r}=(1-\epsilon)(\Gamma-1) c^{2}\frac{dm}{dr}
-(\hat{\gamma}-1)\left(\frac{3}{r}-\frac{1}{\Gamma} \frac{d \Gamma}{d r}\right) U~,
\end{equation}
where ${dm}/{dr}=n_{_{\rm ISM}} m_p r^{2}$
with $n_{_{\rm ISM}}$ being particle density of
interstellar medium (ISM) and $m_p$ being the proton mass,
and $\Gamma(\theta,r)$, $m(\theta,r)$, $U(\theta,r)$, and $\epsilon$
are the bulk Lorentz factor, the sweep-up mass per solid angle,
the internal energy, and the radiation efficiency
of electrons in the external-forward shock, respectively.
The adiabatic index is $\hat{\gamma} \simeq
(5-1.21937\zeta+0.18203 \zeta^{2}
-0.96583 \zeta^{3}+2.32513 \zeta^{4}
-2.39332 \zeta^{5}+1.07136 \zeta^{6})/3$
with $\zeta \equiv \Theta /(0.24+\Theta)$,
$\Theta \simeq
\left(\frac{\Gamma \beta}{3}\right)
\left[\frac{\Gamma \beta+1.07(\Gamma \beta)^{2}}
{1+\Gamma \beta+1.07(\Gamma \beta)^{2}}\right]$,
and $\beta = \sqrt{1-1/\Gamma^2}$
(\citealp{Pe'er_A-2012ApJ.752L.8P}).
The initial Lorentz factor of patch is set as $\Gamma_0(\theta)={E_{\rm iso}(\theta)/[{M_{\rm iso}(\theta) c^2}}]+1$.
Given an appropriate initial value of $U$,
the $r$-dependent $\Gamma$ can be obtained for each patch.
The patches with $\Gamma_0(\theta) < 1.4$ are neglected in our calculations
and not involved in the estimation of baryon loading.
$\epsilon_e$ and $\epsilon_B$ are introduced to represent the fractions of the shock energy
used to accelerate electrons and going into the magnetic energy, respectively.
Then, the magnetic field behind the shock
is $B^{\prime}={(32\pi { \epsilon _B}{n_{{\rm{ISM}}}})^{1/2}}\Gamma c$.
The sweep-up electrons are accelerated
to a power-law distribution of Lorentz factor $\gamma_e$,
i.e., $Q\propto {\gamma '_e}^{ - p}$ for
$\gamma_{e, \min}^{\prime}
\leqslant \gamma_{e} \leqslant \gamma_{e,\max}^{\prime}$,
where $p (> 2)$ is the power-law index,
$\gamma_{e,\min}=\epsilon_e(p-2)m_{\rm p}\Gamma/[(p-1)m_{\rm e}]$ (\citealp{Sari_R-1998-Piran_T-ApJ.497L.17S}),
and $\gamma_{e,\max}=\sqrt {9m_{\rm{e}}^2{c^4}/(8B'q_e^3)}$ with $q_{\rm e}$ being the electron charge (e.g., \citealp{Kumar_P-2012-Hernandez_RA-MNRAS.427L.40K}).
Then, one can have $\epsilon=\epsilon_{\rm rad}\epsilon_e$
with $\epsilon_{\rm rad}={\rm min}\{1,(\gamma_{e,\min}/\gamma_{e,c})^{(p-2)}\}$
(\citealp{Fan_YZ-2006-Piran_T-MNRAS.369.197F}),
where $\gamma_{e,c}=6 \pi m_e c/(\sigma_{\rm T}\Gamma {B^{\prime}}^2 t^{\prime})$
is the efficient cooling Lorentz factor of electrons.

In the X-ray/optical/radio bands,
the main radiation mechanism of the external-forward shock in GRBs
is the synchrotron radiation of the sweep-up electrons
(\citealp{Sari_R-1998-Piran_T-ApJ.497L.17S}; \citealp{Sari_R-1999-Piran_T-ApJ.517L.109S}).
We denote the instantaneous electron spectrum per solid angle at $r$ and $\theta$ as $n'_e(r,\theta,{\gamma'_e})$,
of which the evolution can be solved based on the continuity equation (e.g., \citealp{Liu_K-2020-Lin_DB-ApJ.893L.14L}).
The spectral power of synchrotron radiation of $n'_{\rm e}(r,\theta,{\gamma'_e})$ at a given frequency $\nu'$ is
$P'(\nu',r,\theta)=\frac{\sqrt{3}q_{\rm e}^3B'}{m_{\rm e}c^2}\int_{0}^{\gamma'_{\rm max}}F\left({\nu'}/{\nu'_{\rm c}}\right)n'_ed\gamma'_e,
$
where $F(x)=x\int_{x}^{+\infty}K_{5/3}(k)dk$ with $K_{5/3}(k)$ being the modified Bessel function of 5/3 order
and $\nu'_{\rm c}=3q_{\rm e}B'{\gamma'_e}^2/(4\pi m_{\rm e}c)$.
By summing the flux from each patch observed at a same observer time $t_{\rm obs}$,
the total observed flux can be obtained.

\emph{\textbf{Superluminal Motion of Flux Centroid}\;}
The flux centroid motion, as well as the axial ratio of the image, are powerful tools to constrain the jet structure (\citealp{Gill_R-2018-Granot_J-MNRAS.478.4128G}).
In our calculations,
we record the location $(x,y,z)$=($r\sin \theta \cos \varphi $, $r\sin \theta \sin \varphi $, $r\cos \theta $)
of the maximum flux in the radio image at 3\;GHz band and the observer time $t_{\rm obs}$.
Then, the apparent velocity of the flux centroid
over a time interval $\delta t_{\rm obs}=t_{\rm obs,2}-t_{\rm obs,1}$ is estimated as
$\beta_{\rm app}\equiv
v_{\rm app}/c=
\sqrt{(x_2^{\prime}-x_1^{\prime})^2+(y_2^{\prime}-y_1^{\prime})^2}/(\delta t_{\rm obs} c)$,
where $x' = x\cos {\theta _v} - z\sin {\theta _v}$, $y' = y$,
and $(x_1,y_1,z_1)$ and $(x_2,y_2,z_2)$ are corresponding to the observer time $t_{\rm obs,1}$ and $t_{\rm obs,2}$, respectively.
It should be pointed out that
an elliptical Gaussian fitting of the radio image
should be performed in order to better estimate the location of flux centroid
(e.g., \citealp{Granot_J-2018-De_Colle_F-MNRAS.481.2711G, Lu_WB-2020-Beniamini_P-arXiv200510313L}).
However, our approach based on the location of maximum flux would significantly reduce the computation time.

\section{Fitting Result and Discussion}
\subsection{Fitting on the Afterglows of GRB~170817A}
\begin{figure}
\centering
\includegraphics[width=0.45\textwidth]{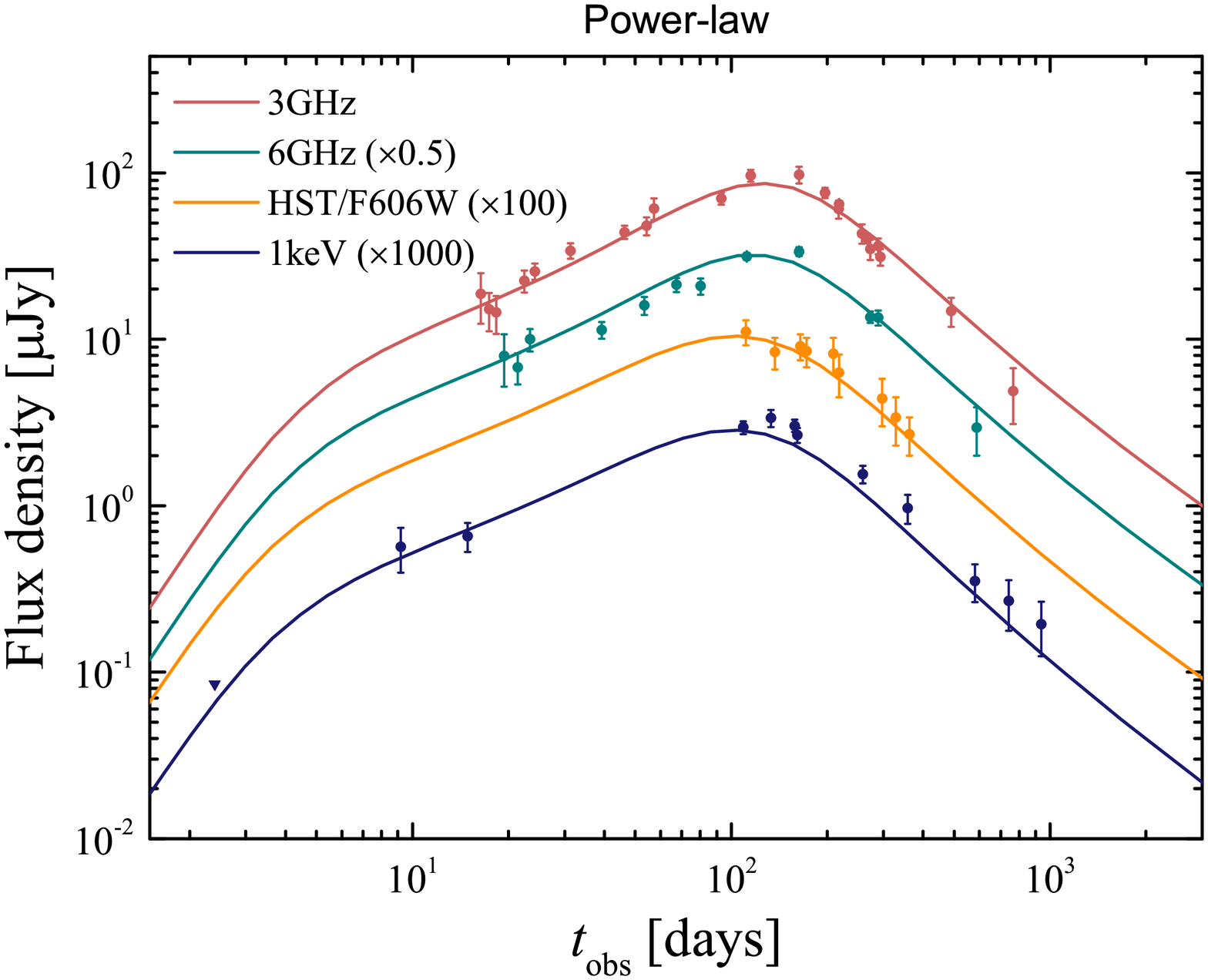}
\includegraphics[width=0.45\textwidth]{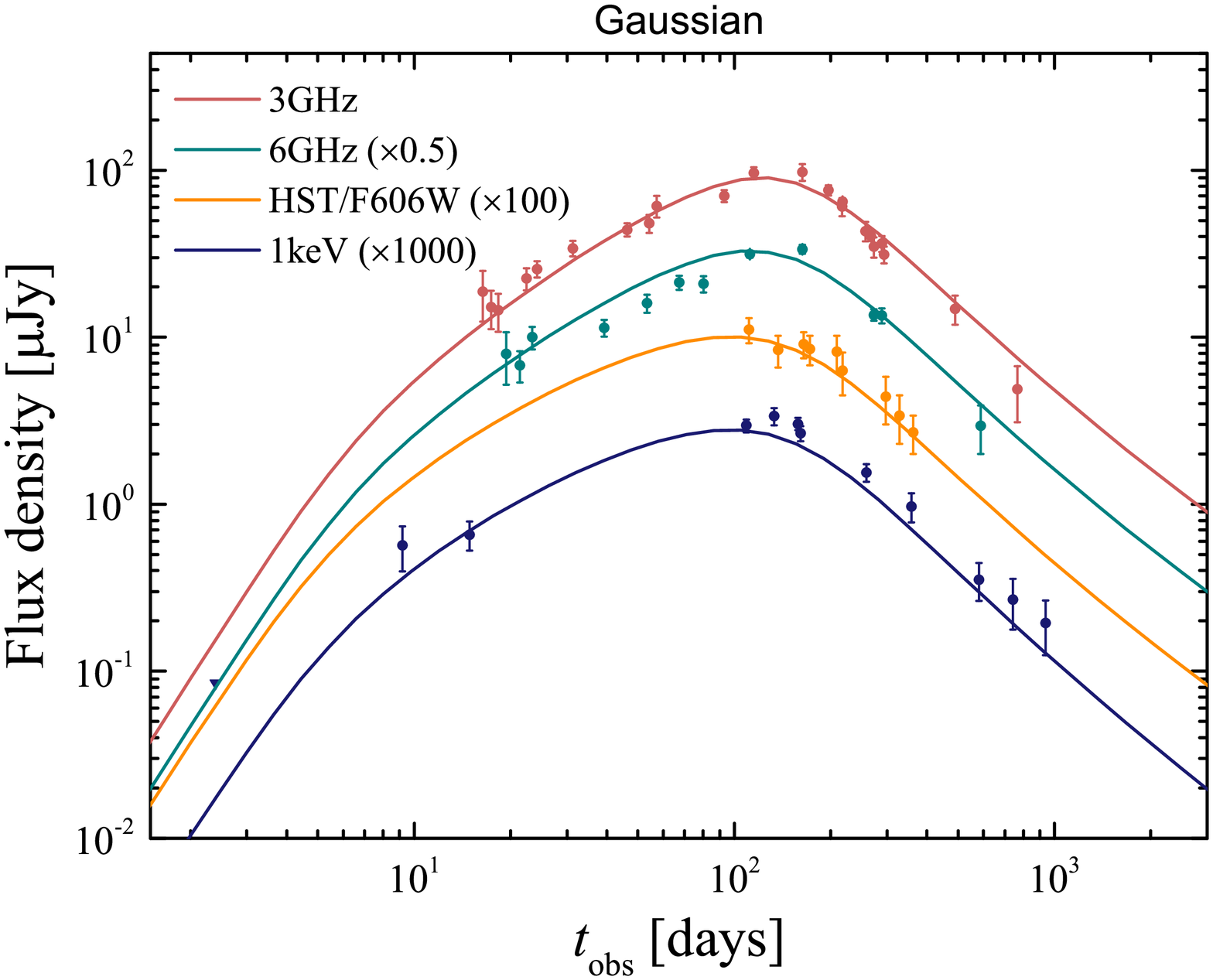}
\caption{Fitting result (solid lines) of afterglows in GRB~170817A.
The observational data are described with circles,
where the triangle is the upper limit of X-ray. }\label{MyFigA}
\end{figure}

\begin{figure}
\centering
\includegraphics[width=0.45\textwidth]{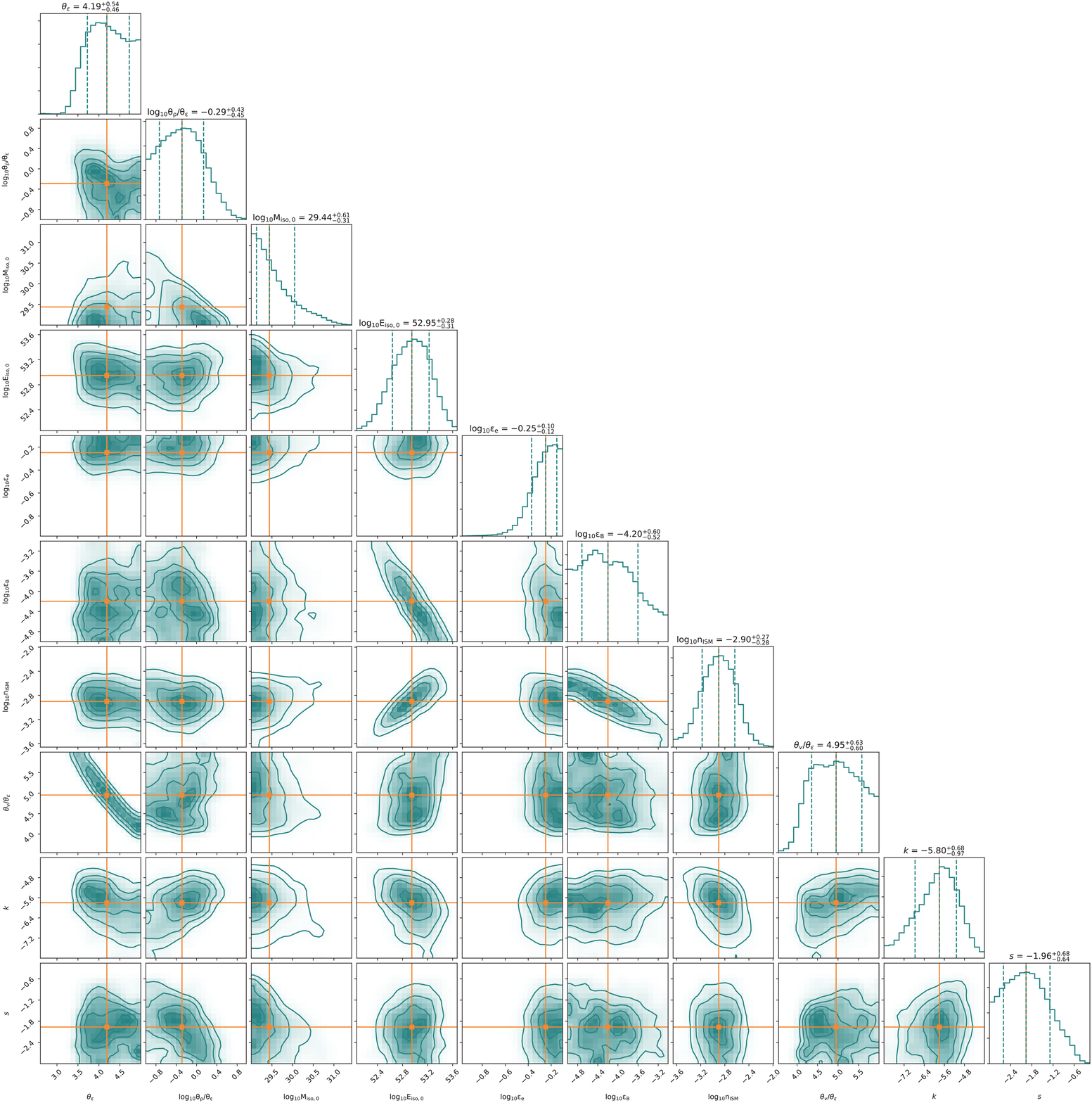}
\includegraphics[width=0.45\textwidth]{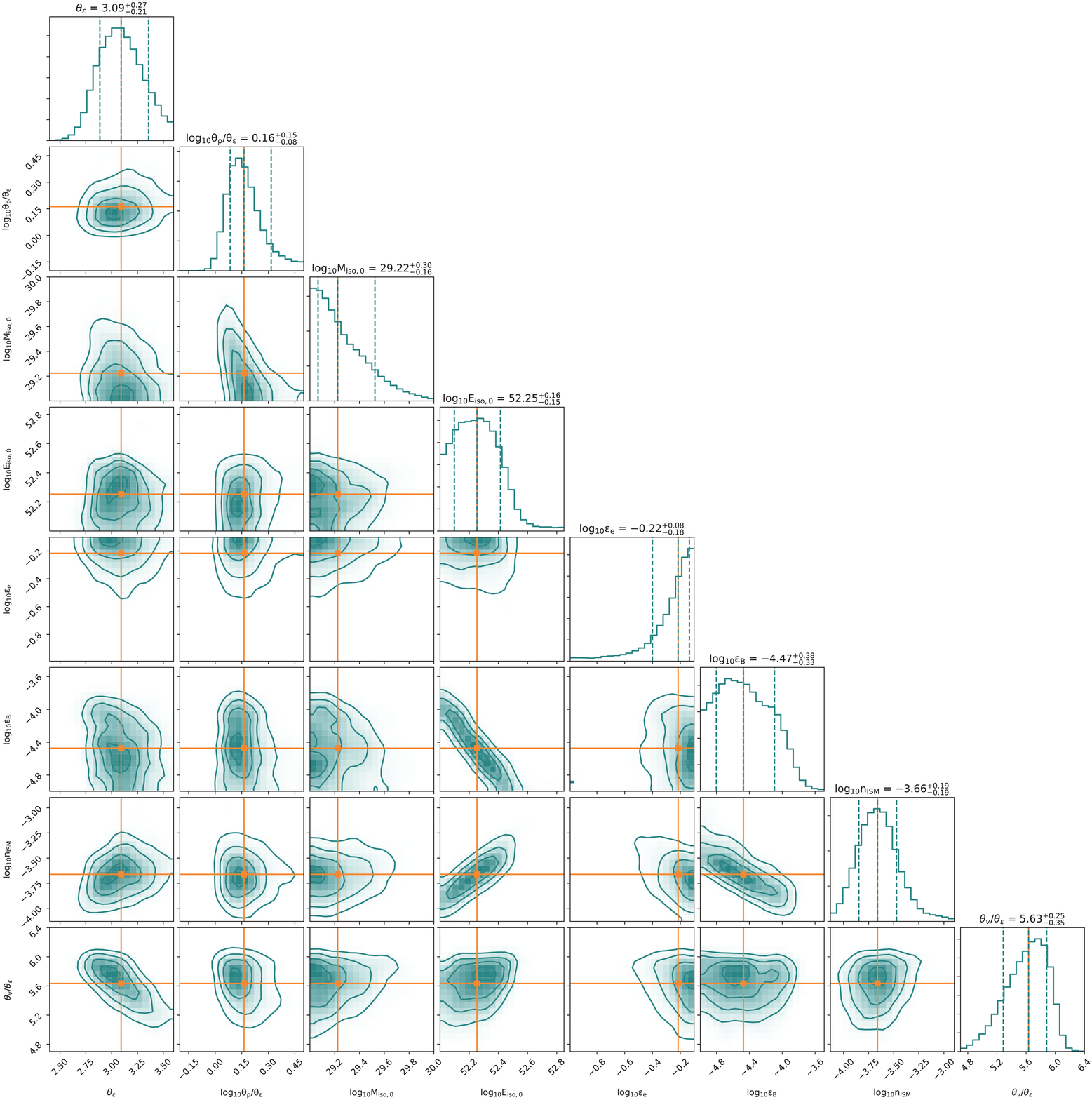}
\caption{Corner plot for the power-law (left) and the Gaussian (right) structured jet model fitting to the afterglows of GRB~170817A.}\label{MyFigB}
\end{figure}

The X-ray/optical/radio afterglows of GRB~170817A is fitted with our structured jet.
Owing to the different process of data processing,
the data from different groups may be a lack of uniformity.
Recently, \cite{Makhathini_S-2020-Mooley_KP-arXiv200602382M}
has compiled and unified all of the observational data about the afterglows of GRB~170817A.
We take their result at 1\;keV, F606W, 6\;GHz, and 3\;GHz bands as our fitting dataset,
which are shown in Figure~{\ref{MyFigA}}.
In addition, the superluminal motion $\beta_{\rm app}=4.1 \pm 0.5$ between
75 and 230 days is also used in our fittings (\citealp{Mooley_KP-2018-Deller_AT-Natur.561.355M}).
Our work includes the fitting with power-law and Gaussian structured jet model.
The fitting is performed based on the Markov Chain Monte Carlo (MCMC) method
to produce posterior predictions for the model parameters.
The Python package {\tt emcee} (\citealp{Foreman-Mackey_D-2013-Hogg_DW-PASP.125.306F})
is used for our MCMC sampling,
where $N_{\rm walkers}\times N_{\rm steps}=25\times2000$ is adopted
and the initial 10\% iterations are used for burn-in.
The projections of the posterior distribution
in 1-D and 2-D for the physical parameters
in the power-law model (i.e., $E_{\rm iso,0}$, $M_{\rm iso,0}$, $\epsilon_e$, $\epsilon_B$, $n_{\rm ISM}$,
$\theta_\varepsilon$, $\theta_\rho/\theta_\varepsilon$, $\theta_{\rm v}/\theta_\varepsilon$, $k$, $s$)
and the Gaussian model (i.e., $E_{\rm iso,0}$, $M_{\rm iso,0}$, $\epsilon_e$, $\epsilon_B$, $n_{\rm ISM}$,
$\theta_\varepsilon$, $\theta_\rho/\theta_\varepsilon$, $\theta_{\rm v}/\theta_\varepsilon$)
are presented in Figure~{\ref{MyFigB}}.
In this work, $p=2.16$ is adopted by considering the spectrum fitting results (e.g., \citealp{Fong_W-2019-Blanchard_PK-ApJ.883L.1F}).
The obtained parameters at the $1\sigma$ confidence level are reported in Table~{\ref{MyTableA}},
and the optimal fitting results are shown in Figure~{\ref{MyFigA}}.

Notice that recent works
(e.g.,
\citealp{Beniamini_P-2020-Granot_J-MNRAS.493.3521B,
Nakar_E-2020-Piran_T-arXiv200501754N, Ryan_G-2020-Eerten_H-ApJ.896..166R})
have pointed out the wide range of $\theta_\varepsilon$ and $\theta_{\rm v}$ values
obtained from afterglows' fittings is due to the lack of key information for fitting.
\cite{Beniamini_P-2020-Granot_J-MNRAS.493.3521B} and
\cite{Nakar_E-2020-Piran_T-arXiv200501754N} clearly indicates that the ratio
$\theta_{\rm v}/\theta_\varepsilon$ is the only quantity
that can be determined from the light curve alone.
Here, we use this ratio as one of the free parameters
and find that $\theta_{\rm v}/\theta_\varepsilon$ has a value $\sim5$ with maximum likelihood under the power-law structured jet model,
which has a high agreement with those works combined with the VLBI superluminal measurement in the fittings
(\citealp{Hotokezaka_K-2019-Nakar_E-NatAs.3.940H, Ghirlanda_G-2019-Salafia_OS-Sci.363.968G}).
The degeneracy between $\theta_\varepsilon$ and $\theta_{\rm v}/\theta_\varepsilon$
in Figure~{\ref{MyFigB}} suggests that
the viewing angle $\theta_{\rm v}$ can be robustly estimated from our fittings.
Our obtained $\theta_\varepsilon$
and $\theta_{\rm v}$
are consistent with the results estimated based on the prompt emission of GRB~170817
(e.g., \citealp{Mooley_KP-2018-Deller_AT-Natur.561.355M,
Beniamini_P-2019-Petropoulou_M-MNRAS.483.840B}).
The ISM number density is well below the upper limit on the density of ionized and neutral particle density
(e.g., \citealp{Hallinan_G-2017-Corsi_A-Sci.358.1579H}, \citealp{Hajela_A-2019-Margutti_R-ApJ.886L.17H}).
The small allowed spread in $\epsilon_e$ is consistent with the results of
\cite{Beniamini_P-2017-van_der_Horst_AJ-MNRAS.472.3161B}.
For our power-law structured jet,
the value of $k\sim -6$ and $s\sim -2$ means $\Gamma(\theta)\propto{{\theta}^{-4}}$ at large angle.
This is consistent with that found in \cite{Ghirlanda_G-2019-Salafia_OS-Sci.363.968G},
of which $E(\theta)\propto{{\theta}^{-5.5}}$
and $\Gamma(\theta)\propto{{\theta}^{-3.5}}$ is found.
\cite{Hotokezaka_K-2019-Nakar_E-NatAs.3.940H} find
$E(\theta)\propto{{\theta}^{-4.5}}$
and $\Gamma(\theta)\propto{{\theta}^{-4.5}}$
by frozen the value of $\Gamma(\theta=0)=600$.
Based on our optimal fitting result,
$\Gamma(\theta=0)=360$ rather than $600$ is obtained.
Then, the low value of power-law index in the $\theta-\Gamma$ relation found in
\cite{Hotokezaka_K-2019-Nakar_E-NatAs.3.940H} may be owing to the high value of $\Gamma(\theta=0)$ adopted in their fittings.

\begin{figure}
\centering
\includegraphics[width=1.0\textwidth]{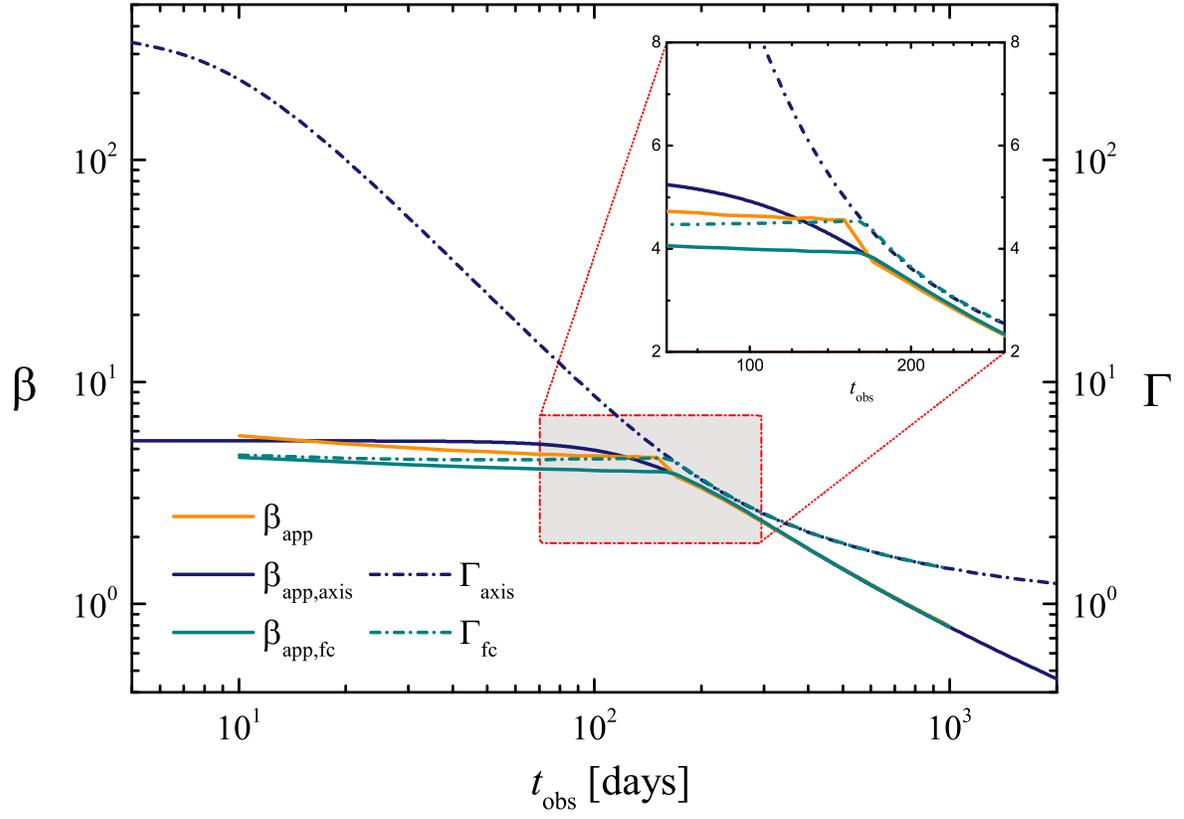}
\caption{Dependence of $\beta _{\rm app}$, $\beta _{\rm app, axis}$, $\beta_{\rm app,fc}$, $\Gamma_{\rm axis}$, and $\Gamma_{\rm fc}$ on $t_{\rm obs}$, where the meanings of lines are shown in the figure.
The inset shows zoomed-in view of the period from 70 days to 300 days.}\label{MyFigC}
\end{figure}

%\subsection{Superluminal Motion from VLBI measurement}\label{VLBI}
In Figure~{\ref{MyFigC}},
we show the evolution of the apparent velocity $\beta_{\rm app}$
of flux centroid motion at 3\;GHz band based on our optimal fitting result of the power-law model,
where the time interval of 10~days is adopted.
For comparison, we also show other noteworthy parameters in this figure:
\textcircled{\footnotesize{1}}
$\Gamma_{\rm axis}$, the bulk Lorentz factor of the jet flow along the jet axis;
\textcircled{\footnotesize{2}}
$\Gamma_{\rm fc}$, the bulk Lorentz factor of flux centroid;
\textcircled{\footnotesize{3}}
${\beta _{{\rm{app}},{\rm{axis}}}} = {\beta _{{\rm{axis}}}}\sin {\theta _{\rm{v}}}/(1 - {\beta _{{\rm{axis}}}}\cos {\theta _{\rm{v}}})$
and
$\beta_{\rm app,fc}={\beta _{\rm fc}}\sin \alpha_{\rm fc}/(1 - \beta _{\rm fc}\cos \alpha_{\rm fc})$,
where
$\beta _{\rm axis} = (1 - \Gamma _{\rm axis}^{ - 2})^{1/2}$,
$\beta _{\rm fc} = ({1 - \Gamma _{\rm fc}^{ - 2}})^{1/2}$,
and $\alpha_{\rm fc}$ is the angle
between the flux centroid and the line of sight (\citealp{Rees_MJ-1966-Natur.211.468R}).
%We plot the value of $\beta_{\rm app}$, $\beta_{\rm app,axis}$, $\beta_{\rm app,fc}$,
%$\Gamma_{\rm axis}$, and $\Gamma_{\rm fc}$
%from $t_{\rm obs}=5$~days to $1000$~days.
In the inset of this figure,
we zoom in the figure for the details in the period of $t_{\rm obs}\sim [70, 300]$~days.
One can find that $\Gamma_{\rm fc}$ is almost a constant
during the early phase ($t_{\rm obs}\lesssim 160$~days),
which is corresponding to the rising phase of afterglows.
It indicates that the flux centroid tracks a region
which has a specific bulk Lorentz factor in this phase.
%As time goes on, a growing number of patches becomes visible
%and thus the flux centroid would move towards the jet axis.
%Once the emission of the jet is dominated by that from the jet core,
%the flux centroid would overlap the jet axis.
%Correspondingly, the $\Gamma_{\rm fc}$ would be the same as
%$\Gamma_{\rm axis}$.
%Thus, the $\beta _{\rm app}$, $\beta _{\rm app, axis}$, and $\beta_{\rm app,fc}$
%would be the same in the later phase,
%as shown in Figure~{\ref{MyFigC}}.
%Besides, $\beta _{\rm app}$, $\beta _{\rm app, axis}$, and $\beta_{\rm app,fc}$
%have slightly differences in the early stage.
At the early stage,
$\beta_{\rm app}$ is always larger than $\beta_{\rm app,fc}$,
which stems from the movement of the flux centroid region in the $\theta$-direction from $\theta\sim \theta_{\rm v}$ to $\theta=0$.
%$\beta_{\rm app, fc}$ always lower than $\beta _{\rm app, axis}$, but $\beta_{\rm app}$ is not,
%especially in the period between 120 and 160~days
%in where $\beta_{\rm app}$ is larger than $\beta_{\rm app, axis}$.
%We notice that the behavior of the above five quantities
%is closely related to the jet structure, viewing angle, and other relevant quantities.
%The widely different behaviors of the above five quantities may appear in the situation with
%different structured jet models, e.g., a Gaussian structured jet, and different viewing angles.
The discrepancy of $\beta_{\rm app}$ with $\beta_{\rm app, fc}$ and ${\beta_{{\rm{app, axis}}}}$
reminds that the apparent velocity of a structured jet should be well estimated in the fittings of its afterglows.

\subsection{Baryon Loading and Constraining on the Jet Launching Time}
\begin{figure}
\centering
\includegraphics[width=1.0\textwidth]{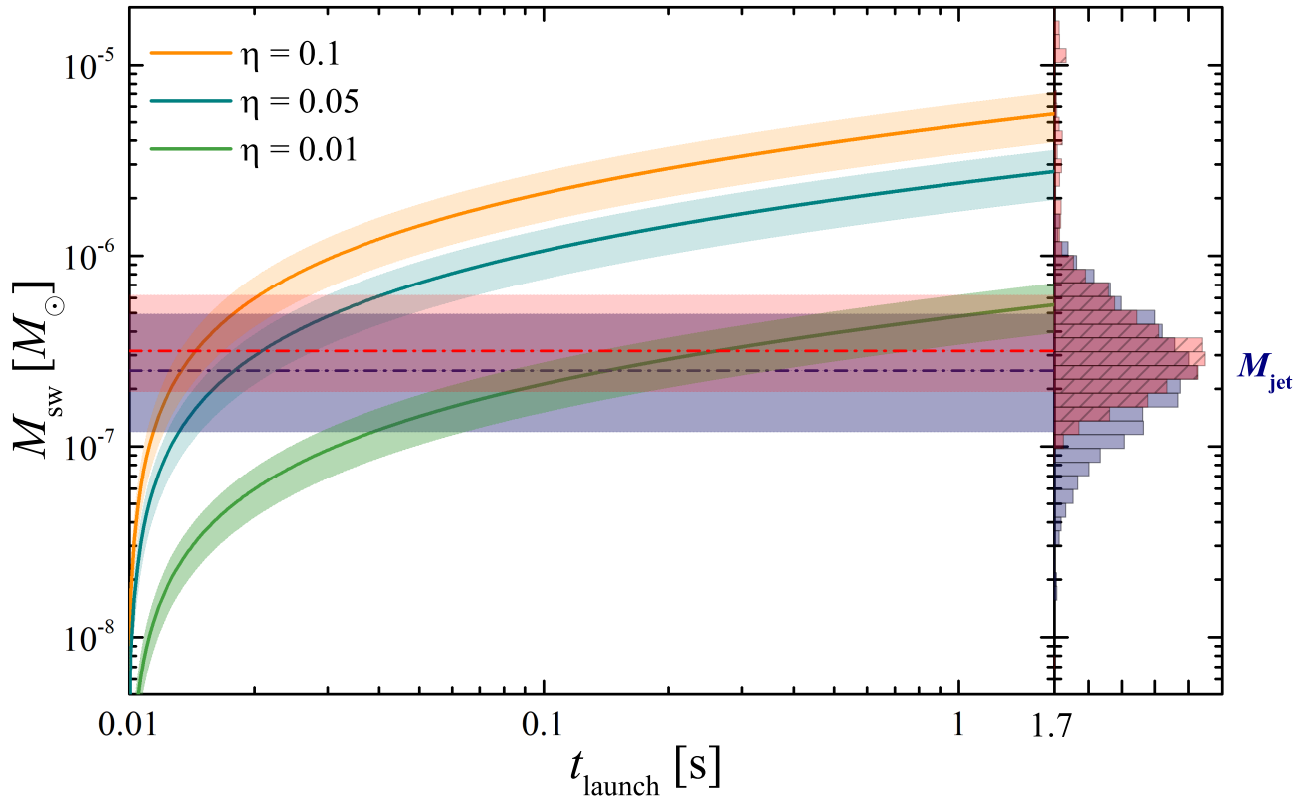}
\caption{Dependence of $M_{\rm sw}$ on $t_{\rm launch}$ (left panel) and
distribution of $M_{\rm jet}$ based on the MCMC samples (right panel).
In the left panel, the yellow, blue, and green lines show the relations of
$M_{\rm sw}-t_{\rm launch}$ with $\eta=$0.1, 0.05, 0.01, respectively.
In addition, the blue (red) dash-dot line shows the median values of $M_{\rm jet}$
by adopting power-law (Gaussian) structured jet in the fitting.
The filled regions around lines display the corresponding $1\sigma$ uncertainties of $M_{\rm sw}$ or $M_{\rm jet}$.
In the right panel, the light-blue and light-red bars are corresponding to the case with power-law
and Gaussian structured jet models, respectively.
}\label{MyFigD}
\end{figure}

According to the MCMC samples,
the baryon loading of the jet $M_{\rm jet}$ can be estimated
and its distribution is shown in the right panel of Figure~{\ref{MyFigD}}.
In this panel, the red and blue histograms are corresponding to those from
the power-law and Gaussian structured jets, respectively.
Then, $M_{\rm jet}=2.49_{-1.30}^{+1.44}\times10^{-7}\;M_{\odot}$
and $M_{\rm jet}=3.16_{-1.23}^{+2.04}\times10^{-7}\;M_{\odot}$
are obtained for the power-law and Gaussian structured jets, respectively.
The median value and $1\sigma$ uncertainty of $M_{\rm jet}$ are also respectively shown
with dash-dotted lines and filled regions in the left panel with the same color in the right panel.

In this subsection,
the obtained $M_{\rm jet}$ is used to constrain the jet launching time
by considering the time/angle-dependent outflow in different ejecting stages.
We assume that when a jet propagates in the surrounding outflowing material,
a fraction $\eta$ of the material in the path of its propagation
is drawn into the jet and becomes a part of $M_{\rm jet}$.
Recent works
which focus on the interaction of a jet with its surrounding ejecta
show that the structure of jet depends strongly on the mixing
taking place both inside the cocoon and along the jet-cocoon interface.
The degree of the mixing would strongly affect the value of $\eta$.
\cite{Gottlieb_O-2020-Bromberg_O-arXiv200711590G, Gottlieb_O-2020-Nakar_E-2020arXiv200602466G},
reveal that the jet power, the angle of the initial injected jet, and the medium density
can strongly influence the degree of mixing.
In this paper, we adopt $\eta=0.1$, 0.05, and 0.01 to represent the degree of mixing
from mild to weak.

By assuming the jet launching time as $t_{\rm launch}$,
the sweep-up mass $M_{\rm sw}$ from outflowing material in the jet can be described as
\begin{eqnarray}\label{Eq:Msw}
M_{\rm jet}\geqslant M_{\rm sw}(t_{\rm launch})\equiv
\int_{0}^{\theta_{\rm ini}}
\eta M_{\rm dyn}(\theta)\sin \theta d\theta
+
2\pi \int_{0}^{\theta_{\rm ini}}\int_{t_{\rm dyn}}^{t_{\rm launch}}
\eta {\dot M_{\rm wind}}(\theta, t^{\prime})
\sin \theta dt^{\prime}d\theta~,
\end{eqnarray}
where $t_{\rm dyn}\sim 10$~ms is the duration of tidal disruption,
$t_{\rm launch}\geqslant t_{\rm dyn}$ is assumed,
and the inequality in the beginning is introduced by involving the initial baryon loading of the jet during its launch.
Here, $M_{\rm dyn}(\theta)=3M_{\rm dyn,tot}\sin^2\theta/(8\pi)$ describes the angle-dependent dynamic ejecta;
${\dot M_{\rm wind}}(\theta ,t)=f {\dot M_{\rm NS}}(t)+3{\dot M_{\rm disk}}(t)\sin^2\theta/(8\pi)$
is the mass in the wind,
${\dot M_{\rm NS}}(t)$
is the quasi-isotropic NS-driven wind,
and
${\dot M_{\rm disk}}(t)$
is the angle-dependent magnetohydrodynamics-viscosity-driven disk wind.
In our work, a uniform mass distribution within the polar angle $\theta_{\rm neu}\lesssim 60^{\circ}$
for the NS-driven wind
is adopted
and thus $f=4\pi /[2\pi (1-\cos \theta_{\rm neu})\times 2]=2$
is introduced to describe the correction of mass distribution in the polar region
for NS-driven wind.
The different parts of ${\dot M_{\rm NS}}(t)={\dot M_{\nu,\rm NS}}(t)+{\dot M_{\rm B, NS}}(t)$ and
${\dot M_{\rm disk}}(t)={\dot M_{\nu,\rm disk}}(t)+{\dot M_{\rm B, disk}}(t)$
are respectively given as
(\citealp{Gill_R-2019-Nathanail_A-ApJ.876.139G})
\begin{eqnarray}\label{nu_NS}
\begin{array}{lll}
{\dot M_{\nu,\rm NS}}(t)
&\approx\left\{\begin{array}{ll}
1.13 \times 10^{-4} (t/{1\;\rm s}) ^{-0.98},\qquad  t_{\mathrm{dyn}}<&t<0.2 \;\mathrm{s}~, \\
 2.02 \times 10^{-4} (t/{1\;\rm s}) ^{-0.62}, \qquad & t \geqslant 0.2 \;\mathrm{s}~,
 \end{array}\right.
\end{array}
\end{eqnarray}
\begin{eqnarray}
{\dot M_{\rm B,NS}}(t)\approx 5.18\times 10^{-3} (t/{1\;\rm s})^{-0.9},\qquad  t \geqslant t_{\rm dyn}~,
\end{eqnarray}
\begin{eqnarray}
{\dot M_{\nu, \rm disk}}(t)&\approx\left\{\begin{array}{ll}
1.43\times 10^{-7} (t/{1\;\rm s})^{-2},  \quad\qquad t_{\rm dyn}< &t<0.12 \;{\rm s}~,\\
1.22\times 10^{-11}(t/{1\;\rm s})^{-6.4},  \qquad        & t \geqslant 0.12 \;{\rm s}~,
\end{array}\right.
\end{eqnarray}
\begin{eqnarray}\label{B_disk}
{\dot M_{\rm B,disk}}(t) \approx 0.012, \qquad  t \geqslant t_{\rm dyn}~,
\end{eqnarray}
in units of $M_{\odot}\;{\rm s}^{-1}$.
The fractional uncertainties of equations (\ref{nu_NS})-(\ref{B_disk})
are introduced in the same way as \cite{Gill_R-2019-Nathanail_A-ApJ.876.139G}.
The description of angle profile of the dynamic ejecta and viscosity-driven winds
in Equation~(\ref{Eq:Msw}) are taken from \cite{Perego_A-2017-Radice_D-ApJ.850L.37P}.
In addition, a series of numerical simulations reported that the total mass of dynamic ejecta
has $M_{\rm dyn,tot} \sim10^{-4}M_{\odot} -10^{-2}M_{\odot}$ (e.g., see \citealp{Nakar_E-2019-arXiv191205659N} for detail)
and $M_{\rm dyn,tot} =(1.5\pm 1.1)\times 10^{-3}M_{\odot}$ is adopted for our analyze
(\citealp{Gill_R-2019-Nathanail_A-ApJ.876.139G}).
The initial jet opening angle $\theta_{\rm ini}$ cannot be known in advance
and the MCMC fitting result
$\theta_{\varepsilon}\sim 4^{\circ}$
is adopted as a possible value of $\theta_{\rm ini}$ in this work.
The jet is assumed non-collimated by the outflows (\citealp{Bromberg_O-2011-Nakar_E-ApJ.740.100B}).

In Figure~{\ref{MyFigD}}, we show the dependence of $M_{\rm sw}$ on $t_{\rm launch}$,
where different value of $\eta$, i.e., $\eta=0.1$, 0.05, 0.01, are adopted.
For a given $M_{\rm jet}$, which is shown with the dash-dotted line and the filled region in Figure~{\ref{MyFigD}},
one can estimate the maximum value of $t_{\rm launch}$ by solving $M_{\rm sw}(t_{\rm launch})=M_{\rm jet}$.
We notice that the waiting time between the merger starting and the jet launching
is less than 0.2~s if $\eta \gtrsim 0.01$ is took.
In addition, $M_{\rm jet}$ needs to increase a factor of two (one magnitude)
in order for the jet launching delay time to increase to around 1~s if $\eta$=0.01 (0.05) is adopted.
A higher $\eta$ will lead to a lower waiting time for the jet launching.
Note that the $M_{\rm sw}$ also depends on
the initial opening angle $\theta_{\rm ini}$,
a higher value of $\theta_{\rm ini}$
means a lower upper limit of launching time, too.

\section{Summary and Conclusion}
In order to infer the baryon loading of the jet,
we use a structured jet to fit the X-ray/optical/radio afterglows of GRB~170817A,
together with the superluminal motion measurement of radio source in this burst.
The structured jet is modelled with angle-dependent energy and baryon loading.
The fitting result of the power-law structured jet
 shows that the ratio between the viewing angle and the jet core angle
is $\theta_{\rm v}/\theta_\varepsilon \sim 5$,
being consistent with other works which involved the superluminal motion measurement in their fittings (\citealp{Hotokezaka_K-2019-Nakar_E-NatAs.3.940H, Ghirlanda_G-2019-Salafia_OS-Sci.363.968G}).
 The obtained $\theta_\varepsilon$  and $\theta_{\rm v}$
are consistent with the results estimated
based on the prompt emission of GRB~170817 (e.g., \citealp{Mooley_KP-2018-Deller_AT-Natur.561.355M, Beniamini_P-2019-Petropoulou_M-MNRAS.483.840B}).
The jet carries the total energy as a few times $10^{49}$~erg.
The on-axis viewed isotropic energy is $10^{52}-10^{53}$~erg,
which is relatively large but still reasonable
(\citealp{Mooley_KP-2018-Deller_AT-Natur.561.355M}).
The ISM number density is
well below the upper limit on the density of the ionized and neutral particles $\sim 10^{-2}~{\rm cm}^{-3}$.
We also studied the motion of the flux centroid in the radio image.
%It is shown that the superluminal motion of a structured jet
%is not directly related to
%the behavior of the jet flow in the jet axis
%but has a more complex behavior.
It should be noted that the behavior of the flux centroid
is wide different for different jet structure and viewing angle.

Based on our fitting result,
the baryon loading of the jet in GRB~170817A is inferred
as $M_{\rm jet}=2.49_{-1.30}^{+1.44}\times10^{-7}\;M_{\odot}$
($M_{\rm jet}=3.16_{-1.23}^{+2.04}\times10^{-7}\;M_{\odot}$) under
the power-law (Gaussian) structured jet model.
By comparing the baryon loading of the jet to the mass outflow in different ejecting stages,
a conservative estimation reveals that the time lag of the jet launch relative to the merger is less than hundreds or tens of milliseconds.
Optimistic estimation would provide a lower upper limit of the jet launching time.
Recently,
works focused on the delay time between the merger and the jet launch
of binary compact star merger have rich conclusions
(e.g., \citealp{Hamidani_H-2020-Kiuchi_K-MNRAS.491.3192H,
Hamidani_H-2020-Ioka_K-arXiv200710690H, Lyutikov_M-2020-MNRAS.491.483L,
Lazzati_D-2019-Perna_R-ApJ.881.89L,Lazzati_D-2020-Ciolfi_R-ApJ.898.59L,
Beniamini_P-2020-Duran_RB-ApJ.895L.33B}).
These works are based on the dynamics of jet during its propagation
in a presupposed isotropic-profile of outflows with or without expanding.
Our discussion is based on a different method, and given an independent constraint.
We notice that the uncertainties in our method mainly
depend on the outflowing rate and angle profile of the merge outflows,
and the fraction of outflowing material drawn into the jet.
The angle profile of outflows
may plays an important role in the collimation of the jet and its breakout from the outflows.
In addition, a successful jet launched by NS
may affect the mass of the outflow within the propagation path of the jet.
Once the mass of the outflow and/or the fraction of outflowing material drawn into the jet
are well established, the launching time of a jet can be
well constrained by utilizing the method proposed in this paper.
The numerical simulations about the propagation of the jet
in the anisotropic and expanding outflows may be necessary
in order to well understand the physics of outflows and jet
formed in the merger
(e.g., \citealp{Murguia-Berthier_A-2020-Ramirez-Ruiz_E-arXiv200712245M}).

The fate of the remnant of GW170817 event is still a mystery
(e.g., \citealp{Granot_J-2017-Guetta_D-ApJ.850L.24G,
Piro_L-2019-Troja_E-MNRAS.483.1912P}).
As \cite{Beniamini_P-2017-Giannios_D-MNRAS.472.3058B,
Beniamini_P-2020-Duran_RB-ApJ.895L.33B} discussed,
the high energy per baryon required for a jet launching delay of $< 0.1$~s
argues against a magnetar central engine for GRB~170817.
Our fitting results of the power-law and the Gaussian structured jet
seems to favor their opinion.
However, the accretion of the disk to the NS may weaken the
constraints of the jet launching time.
Because of the energy released by accretion
might be able to increase the energy per baryon
(e.g., \citealp{Zhang_D-2009-Dai_ZG-ApJ.703.461Z,
Zhang_D-2010-Dai_ZG-ApJ.718.841Z,
Metzger_BD-2018-Thompson_TA-ApJ.856.101M}).
Please see the discussion in \cite{Beniamini_P-2020-Duran_RB-ApJ.895L.33B}.

\acknowledgments
We thank the anonymous referee of this work for useful comments and suggestions that improved the paper.
This work is supported by the National Natural Science Foundation of China (grant Nos.11773007, 11533003, 11851304, U1731239), the Guangxi Science Foundation
(grant Nos. 2018GXNSFFA281010, 2016GXNSFDA380027, 2017AD22006, 2018GXNSFGA281005), the Innovation Team and Outstanding Scholar Program in Guangxi Colleges,
and the Innovation Project of Guangxi Graduate Education (YCSW2019050).
%%%%%%%%%%%%%%%%%%%%%%%%%%%%%%%%%%%%%%%%%%%%%%%%%%%%%%%%%%%%%%%%%%%%%%%%%%%%%%%%%%%%%%%%%%%%%%%%%%%
%%%%%%%%%%%%%%%%%%%%%%%%%%%%%%%%%%%%%%%%%%%%%%%%%%%%%%%%%%%%%%%%%%%%%%%%%%%%%%%%%%%%%%%%%%%%%%%%%%%

\software{\texttt{emcee}
 \citep{Foreman-Mackey_D-2013-Hogg_DW-PASP.125.306F}}

%%%%%%%%%%%%%%%%%%%%%%%%%%%%%%%%%%%%%%%%%%%%%%%%%%%%%%%%%%%%%%%%%%%%%%%%%%%%%%%%%%%%%%%%%%%%%%%%%%%%%%%%%%%%%%%%%%%%%%%%%
\clearpage
\begin{table}[tbph]
\caption{Parameters estimated from the MCMC sampling.}
\centering
\begin{tabular}{lccc}
\hline \hline
Parameter$^{1}$                   &  Power-law                                    &Gaussian                                        & Range        \\                   \hline
 ${\rm log}_{10} E_{\rm iso,0}$ (erg)   &$52.95_{-0.31}^{+0.28}$       &$52.25_{-0.15}^{+0.16}$                             &[52,54]                      \\
 ${\rm log}_{10} M_{\rm iso,0}$ (g)  &$29.44_{-0.31}^{+0.61}$           &$29.22_{-0.16}^{+0.30}$                             &[29,32]   \\
 ${\rm log}_{10} \epsilon_e$    &$-0.25_{-0.12}^{+0.10}$                        &$-0.22_{-0.18}^{+0.08}$                             &[-2,-0.1] \\
 ${\rm log}_{10} \epsilon_B$    &$-4.20_{-0.52}^{+0.60}$                       &$-4.47_{-0.33}^{+0.38}$                              &[-5,-3]                   \\
 ${\rm log}_{10} n_{\rm ISM}$ (${\rm cm}^{-3}$)              &$-2.90_{-0.28}^{+0.27}$   &$-3.66_{-0.19}^{+0.19}$   &[-5,-1]   \\
 $\theta_{\varepsilon}$ (degree)    &$4.19_{-0.46}^{+0.54}$               &$3.09_{-0.21}^{+0.27}$                                 &[2,5]        \\
 ${\rm log}_{10}\theta_{\rho}/\theta_{\varepsilon}$       &$-0.29_{-0.45}^{+0.43}$  &$0.16_{-0.08}^{+0.15}$    &[-1,1]   \\
 $\theta_{\rm v}/\theta_{\varepsilon}$          &$4.95_{-0.60}^{+0.63}$     &$5.63_{-0.35}^{+0.25}$                                    &[2,6] \\
 $k$                        &$-5.80_{-0.97}^{+0.68}$                                                   &--         &[-8,-3]          \\
 $s$                        &$-1.96_{-0.64}^{+0.68}$                                                    &--    &[-3,-0.1]       \\
$p$                          &2.16                       &2.16                            &frozen                        \\
\hline
$\theta_{\rm v}$  (degree)               &$20.65_{-0.66}^{+0.66}$                 &$17.36_{-0.83}^{+0.87}$                           &--   \\
$M_{\rm jet}$ ($M_{\odot}$)      &$2.49_{-1.30}^{+1.44}\times10^{-7}$   &$3.16_{-1.23}^{+2.04}\times10^{-7}$                       &--   \\
$E_{\rm jet}$ (erg)         &$1.25_{-0.56}^{+0.97} \times 10^{49}$                  &$2.71_{-0.90}^{+1.28} \times 10^{49}$         &--      \\
\hline \hline
\end{tabular}\label{MyTableA}%
 \\
 \raggedright
 \footnotesize $^{1}$ There are strong degeneracies between $E_{\rm iso,0}$, $\epsilon_B$, and $n_{\rm ISM}$ as shown in Figure~{\ref{MyFigB}.
 It reveals that a series of parameter sets could fit the afterglows of GRB~170817A
 and the values very far from the estimated median value may be possible.
 In addition, the value of $\theta_{\rm v}$, $M_{\rm jet}$, and $E_{\rm jet}$ are estimated
 based on the MCMC samples rather than the estimated median values.
 }
\end{table}
%%%%%%%%%%%%%%%%%%%%%%%%%%%%%%%%%%%%%%%%%%%%%%%%%%%%%%%%%%%%%%%%%%%%%%%%%%%%%%%%%%%%%%%%%%
%%%%%%%%%%%%%%%%%%%%%%%%%%%%%%%%%%%%%%%%%%%%%%%%%%%%%%%%%%%%%%%%%%%%%%%%%%%%%%%%%%%%%%%%%%

\clearpage
%%%%%%%%%%%%%%%%%%%%%%%%%%%%%%%%%%%%%%%%%%%%%%%%%%%%%%%%%%%%%%%%%%%%%%%%%%%%%%%%%%%%%%%%%%
\clearpage

%%%%%%%%%%%%%%%%%%%%%%%%%%%%%%%%%%%%%%%%%%%%%%%%%%%%%%%%%%%%%%%%%%%%%%%%%%%%%%%%%%%%%%%%%%
\clearpage
%%%%%%%%%%%%%%%%%%%%%%%%%%%%%%%%%%%%%%%%%%%%%%%%%%%%%%%%%%%%%%%%%%%%%%%%%%%%%%%%%%%%%%%%%%
\clearpage

%\bibliographystyle{aasjournal}
%\bibliography{bibliography}

\begin{thebibliography}{}
\expandafter\ifx\csname natexlab\endcsname\relax\def\natexlab#1{#1}\fi
\providecommand{\url}[1]{\href{#1}{#1}}
\providecommand{\dodoi}[1]{doi:~\href{http://doi.org/#1}{\nolinkurl{#1}}}
\providecommand{\doeprint}[1]{\href{http://ascl.net/#1}{\nolinkurl{http://ascl.net/#1}}}
\providecommand{\doarXiv}[1]{\href{https://arxiv.org/abs/#1}{\nolinkurl{https://arxiv.org/abs/#1}}}

\bibitem[{{Abbott} {et~al.}(2017{\natexlab{a}}){Abbott}, {Abbott}, {Abbott},
  {Acernese}, {Ackley}, {Adams}, {Adams}, {Addesso}, {Adhikari}, {Adya}, \&
  et~al.}]{Abbott_BP-2017-Abbott_R-ApJ.850L.39A}
{Abbott}, B.~P., {Abbott}, R., {Abbott}, T.~D., {et~al.} 2017{\natexlab{a}},
  \apjl, 850, L39, \dodoi{10.3847/2041-8213/aa9478}

\bibitem[{{Abbott} {et~al.}(2017{\natexlab{b}}){Abbott}, {Abbott}, {Abbott},
  {Acernese}, {Ackley}, {Adams}, {Adams}, {Addesso}, {Adhikari}, {Adya}, \&
  et~al.}]{Abbott_BP-2017-Abbott_R-ApJ.850L.40A}
---. 2017{\natexlab{b}}, \apjl, 850, L40, \dodoi{10.3847/2041-8213/aa93fc}

\bibitem[{{Abbott} {et~al.}(2017{\natexlab{c}}){Abbott}, {Abbott}, {Abbott},
  {Abernathy}, {Acernese}, {Ackley}, {Adams}, {Adams}, {Addesso}, {Adhikari},
  \& et~al.}]{Abbott_BP-2017-Abbott_R-ApJ.841.89A}
---. 2017{\natexlab{c}}, \apj, 841, 89, \dodoi{10.3847/1538-4357/aa6c47}

\bibitem[{{Abbott} {et~al.}(2017{\natexlab{d}}){Abbott}, {Abbott}, {Abbott},
  {Acernese}, {Ackley}, {Adams}, {Adams}, {Addesso}, {Adhikari}, {Adya}, \&
  et~al.}]{Abbott_BP-2017-Abbott_R-PhRvD.96b2001A}
---. 2017{\natexlab{d}}, \prd, 96, 022001, \dodoi{10.1103/PhysRevD.96.022001}

\bibitem[{{Abbott} {et~al.}(2017{\natexlab{e}}){Abbott}, {Abbott}, {Abbott},
  {Acernese}, {Ackley}, {Adams}, {Adams}, {Addesso}, {Adhikari}, {Adya}, \&
  et~al.}]{Abbott_BP-2017-Abbott_R-ApJ.848L.13A}
---. 2017{\natexlab{e}}, \apjl, 848, L13, \dodoi{10.3847/2041-8213/aa920c}

\bibitem[{{Beniamini} {et~al.}(2020{\natexlab{a}}){Beniamini}, {Duran},
  {Petropoulou}, \& {Giannios}}]{Beniamini_P-2020-Duran_RB-ApJ.895L.33B}
{Beniamini}, P., {Duran}, R.~B., {Petropoulou}, M., \& {Giannios}, D.
  2020{\natexlab{a}}, \apjl, 895, L33, \dodoi{10.3847/2041-8213/ab9223}

\bibitem[{{Beniamini} {et~al.}(2017){Beniamini}, {Giannios}, \&
  {Metzger}}]{Beniamini_P-2017-Giannios_D-MNRAS.472.3058B}
{Beniamini}, P., {Giannios}, D., \& {Metzger}, B.~D. 2017, \mnras, 472, 3058,
  \dodoi{10.1093/mnras/stx2095}

\bibitem[{{Beniamini} {et~al.}(2020{\natexlab{b}}){Beniamini}, {Granot}, \&
  {Gill}}]{Beniamini_P-2020-Granot_J-MNRAS.493.3521B}
{Beniamini}, P., {Granot}, J., \& {Gill}, R. 2020{\natexlab{b}}, \mnras, 493,
  3521, \dodoi{10.1093/mnras/staa538}

\bibitem[{{Beniamini} {et~al.}(2019){Beniamini}, {Petropoulou}, {Barniol
  Duran}, \& {Giannios}}]{Beniamini_P-2019-Petropoulou_M-MNRAS.483.840B}
{Beniamini}, P., {Petropoulou}, M., {Barniol Duran}, R., \& {Giannios}, D.
  2019, \mnras, 483, 840, \dodoi{10.1093/mnras/sty3093}

\bibitem[{{Beniamini} \& {van der
  Horst}(2017)}]{Beniamini_P-2017-van_der_Horst_AJ-MNRAS.472.3161B}
{Beniamini}, P., \& {van der Horst}, A.~J. 2017, \mnras, 472, 3161,
  \dodoi{10.1093/mnras/stx2203}

\bibitem[{{Bromberg} {et~al.}(2011){Bromberg}, {Nakar}, {Piran}, \&
  {Sari}}]{Bromberg_O-2011-Nakar_E-ApJ.740.100B}
{Bromberg}, O., {Nakar}, E., {Piran}, T., \& {Sari}, R. 2011, \apj, 740, 100,
  \dodoi{10.1088/0004-637X/740/2/100}

\bibitem[{{Bromberg} {et~al.}(2018){Bromberg}, {Tchekhovskoy}, {Gottlieb},
  {Nakar}, \& {Piran}}]{Bromberg_O-2018-Tchekhovskoy_A-MNRAS.475.2971B}
{Bromberg}, O., {Tchekhovskoy}, A., {Gottlieb}, O., {Nakar}, E., \& {Piran}, T.
  2018, \mnras, 475, 2971, \dodoi{10.1093/mnras/stx3316}

\bibitem[{{Burns}(2019)}]{Burns_E-2019arXiv190906085B}
{Burns}, E. 2019, arXiv e-prints, arXiv:1909.06085.
\newblock \doarXiv{1909.06085}

\bibitem[{{Connaughton} {et~al.}(2016){Connaughton}, {Burns}, {Goldstein},
  {Blackburn}, {Briggs}, {Zhang}, {Camp}, {Christensen}, {Hui}, {Jenke},
  {Littenberg}, {McEnery}, {Racusin}, {Shawhan}, {Singer}, {Veitch},
  {Wilson-Hodge}, {Bhat}, {Bissaldi}, {Cleveland}, {Fitzpatrick}, {Giles},
  {Gibby}, {von Kienlin}, {Kippen}, {McBreen}, {Mailyan}, {Meegan}, {Paciesas},
  {Preece}, {Roberts}, {Sparke}, {Stanbro}, {Toelge}, \&
  {Veres}}]{Connaughton_V-2016-Burns_E-ApJ.826L.6C}
{Connaughton}, V., {Burns}, E., {Goldstein}, A., {et~al.} 2016, \apjl, 826, L6,
  \dodoi{10.3847/2041-8205/826/1/L6}

\bibitem[{{Connaughton} {et~al.}(2018){Connaughton}, {Burns}, {Goldstein},
  {Blackburn}, {Briggs}, {Christensen}, {Hui}, {Kocevski}, {Littenberg},
  {McEnery}, {Racusin}, {Shawhan}, {Veitch}, {Wilson-Hodge}, {Bhat},
  {Bissaldi}, {Cleveland}, {Giles}, {Gibby}, {von Kienlin}, {Kippen},
  {McBreen}, {Meegan}, {Paciesas}, {Preece}, {Roberts}, {Stanbro}, \&
  {Veres}}]{Connaughton_V-2018-Burns_E-ApJ.853L.9C}
---. 2018, \apjl, 853, L9, \dodoi{10.3847/2041-8213/aaa4f2}

\bibitem[{{Coulter} {et~al.}(2017){Coulter}, {Foley}, {Kilpatrick}, {Drout},
  {Piro}, {Shappee}, {Siebert}, {Simon}, {Ulloa}, {Kasen}, {Madore},
  {Murguia-Berthier}, {Pan}, {Prochaska}, {Ramirez-Ruiz}, {Rest}, \&
  {Rojas-Bravo}}]{Coulter_DA-2017-Foley_RJ-Sci.358.1556C}
{Coulter}, D.~A., {Foley}, R.~J., {Kilpatrick}, C.~D., {et~al.} 2017, Science,
  358, 1556, \dodoi{10.1126/science.aap9811}

\bibitem[{{Cowperthwaite} {et~al.}(2017){Cowperthwaite}, {Berger}, {Villar},
  {Metzger}, {Nicholl}, {Chornock}, {Blanchard}, {Fong}, {Margutti},
  {Soares-Santos}, {Alexander}, {Allam}, {Annis}, {Brout}, {Brown}, {Butler},
  {Chen}, {Diehl}, {Doctor}, {Drout}, {Eftekhari}, {Farr}, {Finley}, {Foley},
  {Frieman}, {Fryer}, {Garc{\'\i}a-Bellido}, {Gill}, {Guillochon}, {Herner},
  {Holz}, {Kasen}, {Kessler}, {Marriner}, {Matheson}, {Neilsen}, {Quataert},
  {Palmese}, {Rest}, {Sako}, {Scolnic}, {Smith}, {Tucker}, {Williams},
  {Balbinot}, {Carlin}, {Cook}, {Durret}, {Li}, {Lopes}, {Louren{\c{c}}o},
  {Marshall}, {Medina}, {Muir}, {Mu{\~n}oz}, {Sauseda}, {Schlegel}, {Secco},
  {Vivas}, {Wester}, {Zenteno}, {Zhang}, {Abbott}, {Banerji}, {Bechtol},
  {Benoit-L{\'e}vy}, {Bertin}, {Buckley-Geer}, {Burke}, {Capozzi}, {Carnero
  Rosell}, {Carrasco Kind}, {Castander}, {Crocce}, {Cunha}, {D'Andrea}, {da
  Costa}, {Davis}, {DePoy}, {Desai}, {Dietrich}, {Drlica-Wagner}, {Eifler},
  {Evrard}, {Fernand ez}, {Flaugher}, {Fosalba}, {Gaztanaga}, {Gerdes},
  {Giannantonio}, {Goldstein}, {Gruen}, {Gruendl}, {Gutierrez}, {Honscheid},
  {Jain}, {James}, {Jeltema}, {Johnson}, {Johnson}, {Kent}, {Krause}, {Kron},
  {Kuehn}, {Nuropatkin}, {Lahav}, {Lima}, {Lin}, {Maia}, {March}, {Martini},
  {McMahon}, {Menanteau}, {Miller}, {Miquel}, {Mohr}, {Neilsen}, {Nichol},
  {Ogando}, {Plazas}, {Roe}, {Romer}, {Roodman}, {Rykoff}, {Sanchez},
  {Scarpine}, {Schindler}, {Schubnell}, {Sevilla-Noarbe}, {Smith}, {Smith},
  {Sobreira}, {Suchyta}, {Swanson}, {Tarle}, {Thomas}, {Thomas}, {Troxel},
  {Vikram}, {Walker}, {Wechsler}, {Weller}, {Yanny}, \&
  {Zuntz}}]{Cowperthwaite_PS-2017-Berger_E-ApJ.848L.17C}
{Cowperthwaite}, P.~S., {Berger}, E., {Villar}, V.~A., {et~al.} 2017, \apjl,
  848, L17, \dodoi{10.3847/2041-8213/aa8fc7}

\bibitem[{{Fan} \& {Piran}(2006)}]{Fan_YZ-2006-Piran_T-MNRAS.369.197F}
{Fan}, Y., \& {Piran}, T. 2006, \mnras, 369, 197,
  \dodoi{10.1111/j.1365-2966.2006.10280.x}

\bibitem[{{Fong} {et~al.}(2019){Fong}, {Blanchard}, {Alexander}, {Strader},
  {Margutti}, {Hajela}, {Villar}, {Wu}, {Ye}, {Berger}, {Chornock},
  {Coppejans}, {Cowperthwaite}, {Eftekhari}, {Giannios}, {Guidorzi},
  {Kathirgamaraju}, {Laskar}, {Macfadyen}, {Metzger}, {Nicholl}, {Paterson},
  {Terreran}, {Sand}, {Sironi}, {Williams}, {Xie}, \&
  {Zrake}}]{Fong_W-2019-Blanchard_PK-ApJ.883L.1F}
{Fong}, W., {Blanchard}, P.~K., {Alexander}, K.~D., {et~al.} 2019, \apjl, 883,
  L1, \dodoi{10.3847/2041-8213/ab3d9e}

\bibitem[{{Foreman-Mackey} {et~al.}(2013){Foreman-Mackey}, {Hogg}, {Lang}, \&
  {Goodman}}]{Foreman-Mackey_D-2013-Hogg_DW-PASP.125.306F}
{Foreman-Mackey}, D., {Hogg}, D.~W., {Lang}, D., \& {Goodman}, J. 2013, \pasp,
  125, 306, \dodoi{10.1086/670067}

\bibitem[{{Ghirlanda} {et~al.}(2019){Ghirlanda}, {Salafia}, {Paragi},
  {Giroletti}, {Yang}, {Marcote}, {Blanchard}, {Agudo}, {An}, {Bernardini},
  {Beswick}, {Branchesi}, {Campana}, {Casadio}, {Chassand e-Mottin}, {Colpi},
  {Covino}, {D'Avanzo}, {D'Elia}, {Frey}, {Gawronski}, {Ghisellini}, {Gurvits},
  {Jonker}, {van Langevelde}, {Melandri}, {Moldon}, {Nava}, {Perego},
  {Perez-Torres}, {Reynolds}, {Salvaterra}, {Tagliaferri}, {Venturi},
  {Vergani}, \& {Zhang}}]{Ghirlanda_G-2019-Salafia_OS-Sci.363.968G}
{Ghirlanda}, G., {Salafia}, O.~S., {Paragi}, Z., {et~al.} 2019, Science, 363,
  968, \dodoi{10.1126/science.aau8815}

\bibitem[{{Gill} \& {Granot}(2018)}]{Gill_R-2018-Granot_J-MNRAS.478.4128G}
{Gill}, R., \& {Granot}, J. 2018, \mnras, 478, 4128,
  \dodoi{10.1093/mnras/sty1214}

\bibitem[{{Gill} {et~al.}(2019){Gill}, {Nathanail}, \&
  {Rezzolla}}]{Gill_R-2019-Nathanail_A-ApJ.876.139G}
{Gill}, R., {Nathanail}, A., \& {Rezzolla}, L. 2019, \apj, 876, 139,
  \dodoi{10.3847/1538-4357/ab16da}

\bibitem[{{Goldstein} {et~al.}(2017){Goldstein}, {Veres}, {Burns}, {Briggs},
  {Hamburg}, {Kocevski}, {Wilson-Hodge}, {Preece}, {Poolakkil}, {Roberts},
  {Hui}, {Connaughton}, {Racusin}, {von Kienlin}, {Dal Canton}, {Christensen},
  {Littenberg}, {Siellez}, {Blackburn}, {Broida}, {Bissaldi}, {Cleveland},
  {Gibby}, {Giles}, {Kippen}, {McBreen}, {McEnery}, {Meegan}, {Paciesas}, \&
  {Stanbro}}]{Goldstein_A-2017-Veres_P-ApJ.848L.14G}
{Goldstein}, A., {Veres}, P., {Burns}, E., {et~al.} 2017, \apjl, 848, L14,
  \dodoi{10.3847/2041-8213/aa8f41}

\bibitem[{{Gottlieb} {et~al.}(2020{\natexlab{a}}){Gottlieb}, {Bromberg},
  {Singh}, \& {Nakar}}]{Gottlieb_O-2020-Bromberg_O-arXiv200711590G}
{Gottlieb}, O., {Bromberg}, O., {Singh}, C.~B., \& {Nakar}, E.
  2020{\natexlab{a}}, arXiv e-prints, arXiv:2007.11590.
\newblock \doarXiv{2007.11590}

\bibitem[{{Gottlieb} {et~al.}(2020{\natexlab{b}}){Gottlieb}, {Nakar}, \&
  {Bromberg}}]{Gottlieb_O-2020-Nakar_E-2020arXiv200602466G}
{Gottlieb}, O., {Nakar}, E., \& {Bromberg}, O. 2020{\natexlab{b}}, arXiv
  e-prints, arXiv:2006.02466.
\newblock \doarXiv{2006.02466}

\bibitem[{{Gottlieb} {et~al.}(2018){Gottlieb}, {Nakar}, {Piran}, \&
  {Hotokezaka}}]{Gottlieb_O-2018-Nakar_E-MNRAS.479.588G}
{Gottlieb}, O., {Nakar}, E., {Piran}, T., \& {Hotokezaka}, K. 2018, \mnras,
  479, 588, \dodoi{10.1093/mnras/sty1462}

\bibitem[{{Granot} {et~al.}(2018){Granot}, {De Colle}, \&
  {Ramirez-Ruiz}}]{Granot_J-2018-De_Colle_F-MNRAS.481.2711G}
{Granot}, J., {De Colle}, F., \& {Ramirez-Ruiz}, E. 2018, \mnras, 481, 2711,
  \dodoi{10.1093/mnras/sty2454}

\bibitem[{{Granot} {et~al.}(2017){Granot}, {Guetta}, \&
  {Gill}}]{Granot_J-2017-Guetta_D-ApJ.850L.24G}
{Granot}, J., {Guetta}, D., \& {Gill}, R. 2017, \apjl, 850, L24,
  \dodoi{10.3847/2041-8213/aa991d}

\bibitem[{{Greiner} {et~al.}(2016){Greiner}, {Burgess}, {Savchenko}, \&
  {Yu}}]{Greiner_J-2016-Burgess_JM-ApJ.827L.38G}
{Greiner}, J., {Burgess}, J.~M., {Savchenko}, V., \& {Yu}, H.-F. 2016, \apjl,
  827, L38, \dodoi{10.3847/2041-8205/827/2/L38}

\bibitem[{{Hajela} {et~al.}(2019){Hajela}, {Margutti}, {Alexander},
  {Kathirgamaraju}, {Baldeschi}, {Guidorzi}, {Giannios}, {Fong}, {Wu},
  {MacFadyen}, {Paggi}, {Berger}, {Blanchard}, {Chornock}, {Coppejans},
  {Cowperthwaite}, {Eftekhari}, {Gomez}, {Hosseinzadeh}, {Laskar}, {Metzger},
  {Nicholl}, {Paterson}, {Radice}, {Sironi}, {Terreran}, {Villar}, {Williams},
  {Xie}, \& {Zrake}}]{Hajela_A-2019-Margutti_R-ApJ.886L.17H}
{Hajela}, A., {Margutti}, R., {Alexander}, K.~D., {et~al.} 2019, \apjl, 886,
  L17, \dodoi{10.3847/2041-8213/ab5226}

\bibitem[{{Hallinan} {et~al.}(2017){Hallinan}, {Corsi}, {Mooley}, {Hotokezaka},
  {Nakar}, {Kasliwal}, {Kaplan}, {Frail}, {Myers}, {Murphy}, {De}, {Dobie},
  {Allison}, {Bannister}, {Bhalerao}, {Chandra}, {Clarke}, {Giacintucci}, {Ho},
  {Horesh}, {Kassim}, {Kulkarni}, {Lenc}, {Lockman}, {Lynch}, {Nichols},
  {Nissanke}, {Palliyaguru}, {Peters}, {Piran}, {Rana}, {Sadler}, \&
  {Singer}}]{Hallinan_G-2017-Corsi_A-Sci.358.1579H}
{Hallinan}, G., {Corsi}, A., {Mooley}, K.~P., {et~al.} 2017, Science, 358,
  1579, \dodoi{10.1126/science.aap9855}

\bibitem[{{Hamidani} \& {Ioka}(2020)}]{Hamidani_H-2020-Ioka_K-arXiv200710690H}
{Hamidani}, H., \& {Ioka}, K. 2020, arXiv e-prints, arXiv:2007.10690.
\newblock \doarXiv{2007.10690}

\bibitem[{{Hamidani} {et~al.}(2020){Hamidani}, {Kiuchi}, \&
  {Ioka}}]{Hamidani_H-2020-Kiuchi_K-MNRAS.491.3192H}
{Hamidani}, H., {Kiuchi}, K., \& {Ioka}, K. 2020, \mnras, 491, 3192,
  \dodoi{10.1093/mnras/stz3231}

\bibitem[{{Hotokezaka} {et~al.}(2019){Hotokezaka}, {Nakar}, {Gottlieb},
  {Nissanke}, {Masuda}, {Hallinan}, {Mooley}, \&
  {Deller}}]{Hotokezaka_K-2019-Nakar_E-NatAs.3.940H}
{Hotokezaka}, K., {Nakar}, E., {Gottlieb}, O., {et~al.} 2019, Nature Astronomy,
  3, 940, \dodoi{10.1038/s41550-019-0820-1}

\bibitem[{{Kasen} {et~al.}(2017){Kasen}, {Metzger}, {Barnes}, {Quataert}, \&
  {Ramirez-Ruiz}}]{Kasen_D-2017-Metzger_B-Natur.551.80K}
{Kasen}, D., {Metzger}, B., {Barnes}, J., {Quataert}, E., \& {Ramirez-Ruiz}, E.
  2017, \nat, 551, 80, \dodoi{10.1038/nature24453}

\bibitem[{{Kumar} {et~al.}(2012){Kumar}, {Hern{\'a}ndez}, {Bo{\v{s}}njak}, \&
  {Barniol Duran}}]{Kumar_P-2012-Hernandez_RA-MNRAS.427L.40K}
{Kumar}, P., {Hern{\'a}ndez}, R.~A., {Bo{\v{s}}njak}, {\v{Z}}., \& {Barniol
  Duran}, R. 2012, \mnras, 427, L40, \dodoi{10.1111/j.1745-3933.2012.01341.x}

\bibitem[{{Lattimer} \&
  {Schramm}(1974)}]{Lattimer_JM-1974-Schramm_DN-ApJ.192L.145L}
{Lattimer}, J.~M., \& {Schramm}, D.~N. 1974, \apjl, 192, L145,
  \dodoi{10.1086/181612}

\bibitem[{{Lattimer} \&
  {Schramm}(1976)}]{Lattimer_JM-1976-Schramm_DN-ApJ.210.549L}
---. 1976, \apj, 210, 549, \dodoi{10.1086/154860}

\bibitem[{{Lazzati} {et~al.}(2020){Lazzati}, {Ciolfi}, \&
  {Perna}}]{Lazzati_D-2020-Ciolfi_R-ApJ.898.59L}
{Lazzati}, D., {Ciolfi}, R., \& {Perna}, R. 2020, \apj, 898, 59,
  \dodoi{10.3847/1538-4357/ab9a44}

\bibitem[{{Lazzati} \& {Perna}(2019)}]{Lazzati_D-2019-Perna_R-ApJ.881.89L}
{Lazzati}, D., \& {Perna}, R. 2019, \apj, 881, 89,
  \dodoi{10.3847/1538-4357/ab2e06}

\bibitem[{{Li} \& {Paczy{\'n}ski}(1998)}]{Li_LX-1998-Paczynski_B-ApJ.507L.59L}
{Li}, L.-X., \& {Paczy{\'n}ski}, B. 1998, \apjl, 507, L59,
  \dodoi{10.1086/311680}

\bibitem[{{Li} {et~al.}(2018){Li}, {Liu}, {Yu}, \&
  {Zhang}}]{Li_SZ-2018-Liu_LD-ApJ.861L.12L}
{Li}, S.-Z., {Liu}, L.-D., {Yu}, Y.-W., \& {Zhang}, B. 2018, \apjl, 861, L12,
  \dodoi{10.3847/2041-8213/aace61}

\bibitem[{{Lin} {et~al.}(2018){Lin}, {Liu}, {Lin}, {Wang}, {Gu}, \&
  {Liang}}]{Lin_DB-2018-Liu_T-ApJ.856.90L}
{Lin}, D.-B., {Liu}, T., {Lin}, J., {et~al.} 2018, \apj, 856, 90,
  \dodoi{10.3847/1538-4357/aab3d7}

\bibitem[{{Liu} {et~al.}(2020){Liu}, {Lin}, {Wang}, {Zhou}, {Wang}, \&
  {Liang}}]{Liu_K-2020-Lin_DB-ApJ.893L.14L}
{Liu}, K., {Lin}, D.-B., {Wang}, K., {et~al.} 2020, \apjl, 893, L14,
  \dodoi{10.3847/2041-8213/ab838e}

\bibitem[{{Lu} {et~al.}(2020){Lu}, {Beniamini}, \&
  {McDowell}}]{Lu_WB-2020-Beniamini_P-arXiv200510313L}
{Lu}, W., {Beniamini}, P., \& {McDowell}, A. 2020, arXiv e-prints,
  arXiv:2005.10313.
\newblock \doarXiv{2005.10313}

\bibitem[{{Lyutikov}(2020)}]{Lyutikov_M-2020-MNRAS.491.483L}
{Lyutikov}, M. 2020, \mnras, 491, 483, \dodoi{10.1093/mnras/stz3044}

\bibitem[{{Makhathini} {et~al.}(2020){Makhathini}, {Mooley}, {Brightman},
  {Hotokezaka}, {Nayana}, {Intema}, {Dobie}, {Lenc}, {Perley}, {Fremling},
  {Moldon}, {Lazzati}, {Kaplan}, {Balasubramanian}, {Brown}, {Carbone},
  {Chandra}, {Corsi}, {Camilo}, {Deller}, {Frail}, {Murphy}, {Murphy}, {Nakar},
  {Smirnov}, {Beswick}, {Fender}, {Hallinan}, {Heywood}, {Kasliwal}, {Lee},
  {Lu}, {Rana}, {Perkins}, {White}, {Jozsa}, {Hugo}, \&
  {Kamphuis}}]{Makhathini_S-2020-Mooley_KP-arXiv200602382M}
{Makhathini}, S., {Mooley}, K.~P., {Brightman}, M., {et~al.} 2020, arXiv
  e-prints, arXiv:2006.02382.
\newblock \doarXiv{2006.02382}

\bibitem[{{Matsumoto} {et~al.}(2018){Matsumoto}, {Ioka}, {Kisaka}, \&
  {Nakar}}]{Matsumoto_T-2018-Ioka_K-ApJ.861.55M}
{Matsumoto}, T., {Ioka}, K., {Kisaka}, S., \& {Nakar}, E. 2018, \apj, 861, 55,
  \dodoi{10.3847/1538-4357/aac4a8}

\bibitem[{{Metzger}(2017)}]{Metzger_BD-2017LRR.20.3M}
{Metzger}, B.~D. 2017, Living Reviews in Relativity, 20, 3,
  \dodoi{10.1007/s41114-017-0006-z}

\bibitem[{{Metzger} {et~al.}(2018){Metzger}, {Thompson}, \&
  {Quataert}}]{Metzger_BD-2018-Thompson_TA-ApJ.856.101M}
{Metzger}, B.~D., {Thompson}, T.~A., \& {Quataert}, E. 2018, \apj, 856, 101,
  \dodoi{10.3847/1538-4357/aab095}

\bibitem[{{Mooley} {et~al.}(2018){Mooley}, {Deller}, {Gottlieb}, {Nakar},
  {Hallinan}, {Bourke}, {Frail}, {Horesh}, {Corsi}, \&
  {Hotokezaka}}]{Mooley_KP-2018-Deller_AT-Natur.561.355M}
{Mooley}, K.~P., {Deller}, A.~T., {Gottlieb}, O., {et~al.} 2018, \nat, 561,
  355, \dodoi{10.1038/s41586-018-0486-3}

\bibitem[{{Murguia-Berthier} {et~al.}(2020){Murguia-Berthier}, {Ramirez-Ruiz},
  {De Colle}, {Janiuk}, {Rosswog}, \&
  {Lee}}]{Murguia-Berthier_A-2020-Ramirez-Ruiz_E-arXiv200712245M}
{Murguia-Berthier}, A., {Ramirez-Ruiz}, E., {De Colle}, F., {et~al.} 2020,
  arXiv e-prints, arXiv:2007.12245.
\newblock \doarXiv{2007.12245}

\bibitem[{{Nakar}(2019)}]{Nakar_E-2019-arXiv191205659N}
{Nakar}, E. 2019, arXiv e-prints, arXiv:1912.05659.
\newblock \doarXiv{1912.05659}

\bibitem[{{Nakar} \& {Piran}(2020)}]{Nakar_E-2020-Piran_T-arXiv200501754N}
{Nakar}, E., \& {Piran}, T. 2020, arXiv e-prints, arXiv:2005.01754.
\newblock \doarXiv{2005.01754}

\bibitem[{{Pe'er}(2012)}]{Pe'er_A-2012ApJ.752L.8P}
{Pe'er}, A. 2012, \apjl, 752, L8, \dodoi{10.1088/2041-8205/752/1/L8}

\bibitem[{{Perego} {et~al.}(2017){Perego}, {Radice}, \&
  {Bernuzzi}}]{Perego_A-2017-Radice_D-ApJ.850L.37P}
{Perego}, A., {Radice}, D., \& {Bernuzzi}, S. 2017, \apjl, 850, L37,
  \dodoi{10.3847/2041-8213/aa9ab9}

\bibitem[{{Piro} {et~al.}(2019){Piro}, {Troja}, {Zhang}, {Ryan}, {van Eerten},
  {Ricci}, {Wieringa}, {Tiengo}, {Butler}, {Cenko}, {Fox}, {Khandrika},
  {Novara}, {Rossi}, \& {Sakamoto}}]{Piro_L-2019-Troja_E-MNRAS.483.1912P}
{Piro}, L., {Troja}, E., {Zhang}, B., {et~al.} 2019, \mnras, 483, 1912,
  \dodoi{10.1093/mnras/sty3047}

\bibitem[{{Rees}(1966)}]{Rees_MJ-1966-Natur.211.468R}
{Rees}, M.~J. 1966, \nat, 211, 468, \dodoi{10.1038/211468a0}

\bibitem[{{Ren} {et~al.}(2019){Ren}, {Lin}, {Zhang}, {Li}, {Liu}, {Lu}, {Wang},
  \& {Liang}}]{Ren_J-2019-Lin_DB-ApJ.885.60R}
{Ren}, J., {Lin}, D.-B., {Zhang}, L.-L., {et~al.} 2019, \apj, 885, 60,
  \dodoi{10.3847/1538-4357/ab4188}

\bibitem[{{Ryan} {et~al.}(2020){Ryan}, {Eerten}, {Piro}, \&
  {Troja}}]{Ryan_G-2020-Eerten_H-ApJ.896..166R}
{Ryan}, G., {Eerten}, H.~v., {Piro}, L., \& {Troja}, E. 2020, \apj, 896, 166,
  \dodoi{10.3847/1538-4357/ab93cf}

\bibitem[{{Sari} \& {Piran}(1999)}]{Sari_R-1999-Piran_T-ApJ.517L.109S}
{Sari}, R., \& {Piran}, T. 1999, \apjl, 517, L109, \dodoi{10.1086/312039}

\bibitem[{{Sari} {et~al.}(1998){Sari}, {Piran}, \&
  {Narayan}}]{Sari_R-1998-Piran_T-ApJ.497L.17S}
{Sari}, R., {Piran}, T., \& {Narayan}, R. 1998, \apjl, 497, L17,
  \dodoi{10.1086/311269}

\bibitem[{{Symbalisty} \&
  {Schramm}(1982)}]{Symbalisty_E-1982-Schramm_DN-ApL.22.143S}
{Symbalisty}, E., \& {Schramm}, D.~N. 1982, \aplett, 22, 143

\bibitem[{{Troja} {et~al.}(2019){Troja}, {van Eerten}, {Ryan}, {Ricci},
  {Burgess}, {Wieringa}, {Piro}, {Cenko}, \&
  {Sakamoto}}]{Troja_E-2019-vanEerten_H-MNRAS.tmp.2169T}
{Troja}, E., {van Eerten}, H., {Ryan}, G., {et~al.} 2019, \mnras, 2169,
  \dodoi{10.1093/mnras/stz2248}

\bibitem[{{Villar} {et~al.}(2017){Villar}, {Guillochon}, {Berger}, {Metzger},
  {Cowperthwaite}, {Nicholl}, {Alexander}, {Blanchard}, {Chornock},
  {Eftekhari}, {Fong}, {Margutti}, \&
  {Williams}}]{Villar_VA-2017-Guillochon_J-ApJ.851L.21V}
{Villar}, V.~A., {Guillochon}, J., {Berger}, E., {et~al.} 2017, \apjl, 851,
  L21, \dodoi{10.3847/2041-8213/aa9c84}

\bibitem[{{Waxman} {et~al.}(2018){Waxman}, {Ofek}, {Kushnir}, \&
  {Gal-Yam}}]{Waxman_E-2018-Ofek_EO-MNRAS.481.3423W}
{Waxman}, E., {Ofek}, E.~O., {Kushnir}, D., \& {Gal-Yam}, A. 2018, \mnras, 481,
  3423, \dodoi{10.1093/mnras/sty2441}

\bibitem[{{Yang} {et~al.}(2019){Yang}, {Zhong}, {Zhang}, {Wu}, {Zhang}, {Yang},
  {Cao}, {Gao}, {Zou}, {Wang}, {L{\"u}}, {Cang}, \&
  {Dai}}]{Yang_YS-2019-Zhong_SQ-arXiv191200375Y}
{Yang}, Y.-S., {Zhong}, S.-Q., {Zhang}, B.-B., {et~al.} 2019, arXiv e-prints,
  arXiv:1912.00375.
\newblock \doarXiv{1912.00375}

\bibitem[{{Yu} {et~al.}(2018){Yu}, {Liu}, \&
  {Dai}}]{Yu_YW-2018-Liu_LD-ApJ.861..114Y}
{Yu}, Y.-W., {Liu}, L.-D., \& {Dai}, Z.-G. 2018, \apj, 861, 114,
  \dodoi{10.3847/1538-4357/aac6e5}

\bibitem[{{Zhang}(2018)}]{Zhang_B-2018-pgrb.book.....Z}
{Zhang}, B. 2018, {The Physics of Gamma-Ray Bursts},
  \dodoi{10.1017/9781139226530}

\bibitem[{{Zhang}(2019)}]{Zhang_B-2019-FrPhy.1464402Z}
---. 2019, Frontiers of Physics, 14, 64402, \dodoi{10.1007/s11467-019-0913-4}

\bibitem[{{Zhang} {et~al.}(2018){Zhang}, {Zhang}, {Sun}, {Lei}, {Gao}, {Li},
  {Shao}, {Zhao}, {Hu}, {L{\"u}}, {Wu}, {Fan}, {Wang}, {Castro-Tirado},
  {Zhang}, {Yu}, {Cao}, \& {Liang}}]{Zhang_BB-2018-Zhang_B-NatCo.9.447Z}
{Zhang}, B.-B., {Zhang}, B., {Sun}, H., {et~al.} 2018, Nature Communications,
  9, 447, \dodoi{10.1038/s41467-018-02847-3}

\bibitem[{{Zhang} \& {Dai}(2009)}]{Zhang_D-2009-Dai_ZG-ApJ.703.461Z}
{Zhang}, D., \& {Dai}, Z.~G. 2009, \apj, 703, 461,
  \dodoi{10.1088/0004-637X/703/1/461}

\bibitem[{{Zhang} \& {Dai}(2010)}]{Zhang_D-2010-Dai_ZG-ApJ.718.841Z}
---. 2010, \apj, 718, 841, \dodoi{10.1088/0004-637X/718/2/841}

\bibitem[{{Zhu} {et~al.}(2020){Zhu}, {Yang}, {Liu}, {Huang}, {Zhang}, {Li},
  {Yu}, \& {Gao}}]{Zhu_JP-2020-Yang_YP-ApJ.897.20Z}
{Zhu}, J.-P., {Yang}, Y.-P., {Liu}, L.-D., {et~al.} 2020, \apj, 897, 20,
  \dodoi{10.3847/1538-4357/ab93bf}

\end{thebibliography}

\end{document}